\documentclass[]{spie}  %>>> use for US letter paper
\usepackage{amsmath,amsfonts,amssymb}
\usepackage{subcaption}
\usepackage[skip=5pt]{caption}
\usepackage{graphicx}
\usepackage{float}
\usepackage{fancyhdr}
\usepackage[colorlinks=true, allcolors=blue]{hyperref}

\title{	Overview of the spectrometer optical fiber feed for the Habitable-zone Planet Finder}

\author[a,b]{Shubham Kanodia}
\author[a,b]{Suvrath Mahadevan}
\author[a,b]{Lawrence. W. Ramsey}
\author[a,b,c]{Gudmundur K. Stefansson}
\author[a,b]{Andrew J. Monson}
\author[a,b]{Frederick R. Hearty}
\author[a,b]{Scott Blakeslee}
\author[a,b]{Emily Lubar}
\author[d]{Chad F. Bender}
\author[a,b]{J. P. Ninan}
\author[a,b]{David Sterner}
\author[e,f]{Arpita Roy}
\author[g,h]{Samuel P. Halverson}
\author[i]{Paul M. Robertson}

\affil[a]{Department of Astronomy \& Astrophysics, The Pennsylvania State University, 525 Davey Lab, University Park, PA, USA, 16802}
\affil[b]{Center for Exoplanets \& Habitable Worlds, The Pennsylvania State University, University
Park, PA, USA, 16802}
\affil[c]{NASA Earth and Space Science Fellow}
\affil[d]{Steward Observatory, University of  Arizona, Tucson, AZ, USA, 85721}
\affil[e]{Robert A. Millikan Prize Postdoctoral Fellow}
\affil[f]{Department of Astronomy, California Institute of Technology, Pasadena, CA, USA, 91125}
\affil[g]{Department of Physics \& Astronomy, University of Pennsylvania, 209 S 33rd Street, Philadelphia, PA , USA, 19104}
\affil[h]{NASA Sagan Fellow}
\affil[i]{Department of Physics \& Astronomy, University of California - Irvine, 4129 Frederick Reines Hall, Irvine, CA, USA, 92697}

\authorinfo{Further author information: E-mail: szk381@psu.edu}

\fancypagestyle{spiefoot}
{	
   \fancyhf{}
   \fancyfoot[C]{\footnotesize Proc. of SPIE Vol. 10702 107026Q-\thepage}
}
 
\begin{document} 
\pagestyle{spiefoot}
\maketitle

\begin{abstract}
The Habitable-zone Planet Finder (HPF) is a highly stabilized fiber fed precision radial velocity (RV) spectrograph working in the Near Infrared (NIR):  810 -- 1280 nm . In this paper we present an overview of the preparation of the optical fibers for HPF. The entire fiber train from the telescope focus down to the cryostat is detailed. We also discuss the fiber polishing, splicing and its integration into the instrument using a fused silica puck. HPF was designed to be able to operate in two modes, High Resolution (HR- the only mode mode currently commissioned) and High Efficiency (HE). We discuss these fiber heads and the procedure we adopted to attach the slit on to the HR fibers.
\end{abstract}


\section{INTRODUCTION}
\label{sec:intro} 
% OF introduced in 1970s were recognized for their ability to scramble the light
Optical fibers were introduced in astronomy in the late 1970s and recognized for their multiplexing and scrambling ability\cite{k._serkowski_fabry-perot_1979}. They were proposed as a means of mitigating the guiding errors \cite{black_assessment_1980-1}. Other drivers for using Optical fibers were that it allowed the instrument to be decoupled from the telescope focus and instead to be located in the basement of the observatory. This enabled a constant gravity vector on the instrument and also allows for environmental control\cite{barden_evaluation_1981,heacox_application_1986}. Around the same time, spectroscopic methods were discussed to find extra - solar planets\cite{black_assessment_1980}. 

The first exoplanet around a hydrogen - burning star was discovered in 1995 using the ELODIE spectrometer \cite{mayor_jupiter-mass_1995}. ELODIE was one of the first precision radial velocity (PRV) spectrometers to use an optical fiber feed \cite{baranne_elodie:_1996}. It used a dual fiber feed for simultaneous science and calibration light. This method enabled it to reach an unprecedented RV precision of $\sim$ 13 m/s. This was taken to the next level in 2003 by High Accuracy Radial velocity Planetary Searcher (HARPS)\cite{mayor_setting_2003}. The instrument was housed in a cryostat maintained under vacuum, which helped it achieve exquisite environmental stability. It was one of the first instruments to approach $\sim$ 1 m/s in its RV precision. 

Since then, it has continued to remain the gold standard in RV precision, and has even demonstrated sub m/s precision\cite{lovis_exoplanet_2006}. However, the next generation of optical PRV are aiming for $\sim$ 10 cm/s precision, as that is the magnitude of the Doppler reflex motion of a Sun like star due to a true Earth analogue. Similarly, near-infrared (NIR) instruments are seeking to find an Earth like planet in the Habitable zone\cite{kasting_habitable_1993} around M dwarfs. Since these stars are much smaller and cooler than the Sun, such planets produce a reflex motion of the order of 1 m/s in the NIR. 

HPF is one of these new NIR instruments. It spans 810 to 1280 nm and uses an R4 echelle grating in a white pupil design\cite{mahadevan_habitable-zone_2012}. Deployed at the 10 m class Hobby-Eberly Telescope (HET) in October 2017, it started shared risk science in May 2018. It uses a Teledyne Hawaii-2RG (H2RG) detector with a 1.7 $\mu$m cut off, and the optical bench and optics are cooled down to 180K and kept stable at that temperature using a sophisticated environmental stability system\cite{stefansson_versatile_2016}.

\section{Overview of fiber train for HPF}

\begin{figure}[h]
\centering
\includegraphics[width=0.7\textwidth]{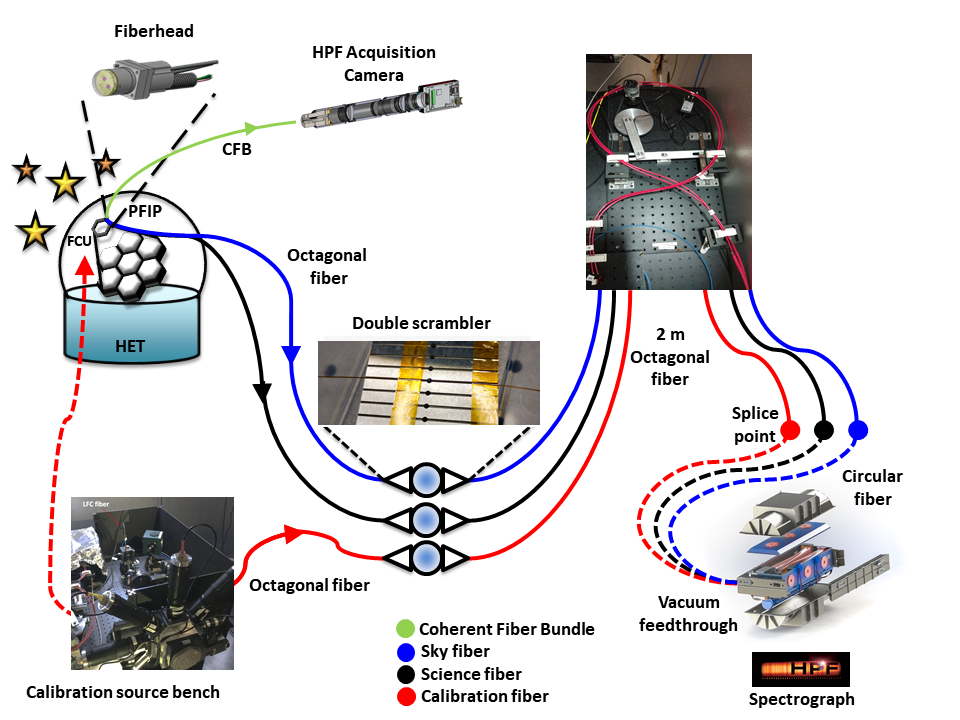}
\caption{\label{fig:schematic} HPF fiber train. From telescope to the Instrument.}
\end{figure}

The fixed elevation and roving pupil design of the Hobby Eberly Telescope leads to illumination variation of the target across a track. This necessitates a stable illumination source which scrambles this variation. Three circular fibers go in to the instrument through vacuum feedthroughs. These are for the science light, sky light and calibration. Figure \ref{fig:schematic} shows the schematic of the fiber train for HPF. Here we discuss the entire High Resolution (HR) mode fiber train. The High Efficiency (HE) mode is exactly the same, and uses the same fibers: the only difference between the two modes is that the HR mode puck inside the cryostat has a nickel slit glued on to it, whereas the HE mode does not. 
The telescope prime-focus end will be discussed in Section $\S$ \ref{sec:telescope_end}.  The octagonal science and sky fiber feed light from the telescope on to the double scrambler. The dimensions of the circular fibers in $\mu$m are 312, 375, 411 (core, cladding, buffer); whereas those for the octagonal are 299, 375, 415  (core, cladding, buffer {these numbers are not exact}). The rationale for using this fiber diameter is explained in \cite{roy_scrambling_2014}.

\subsection{Double scrambler}
 This takes place for the science and sky octagonal fibers from the fiberhead, and the octagonal calibration fiber from the cal bench.  As discussed in \cite{halverson_efficient_2015}, we found the optimum scrambling arrangement to be octagonal + ball lens + octagonal + circular fiber. The input end of the double scrambler consists of the octagonal fibers butt coupled to an AR coated ball lens on to a v-groove block. On the other side is coupled a 2 m patch of octagonal fiber, which is enclosed in the same jacketing as the cryostat fiber. This octagonal patch is then spliced on to the circular cryostat fiber.

\begin{figure}[h]
\centering
\includegraphics[angle = 270, width=0.65\textwidth]{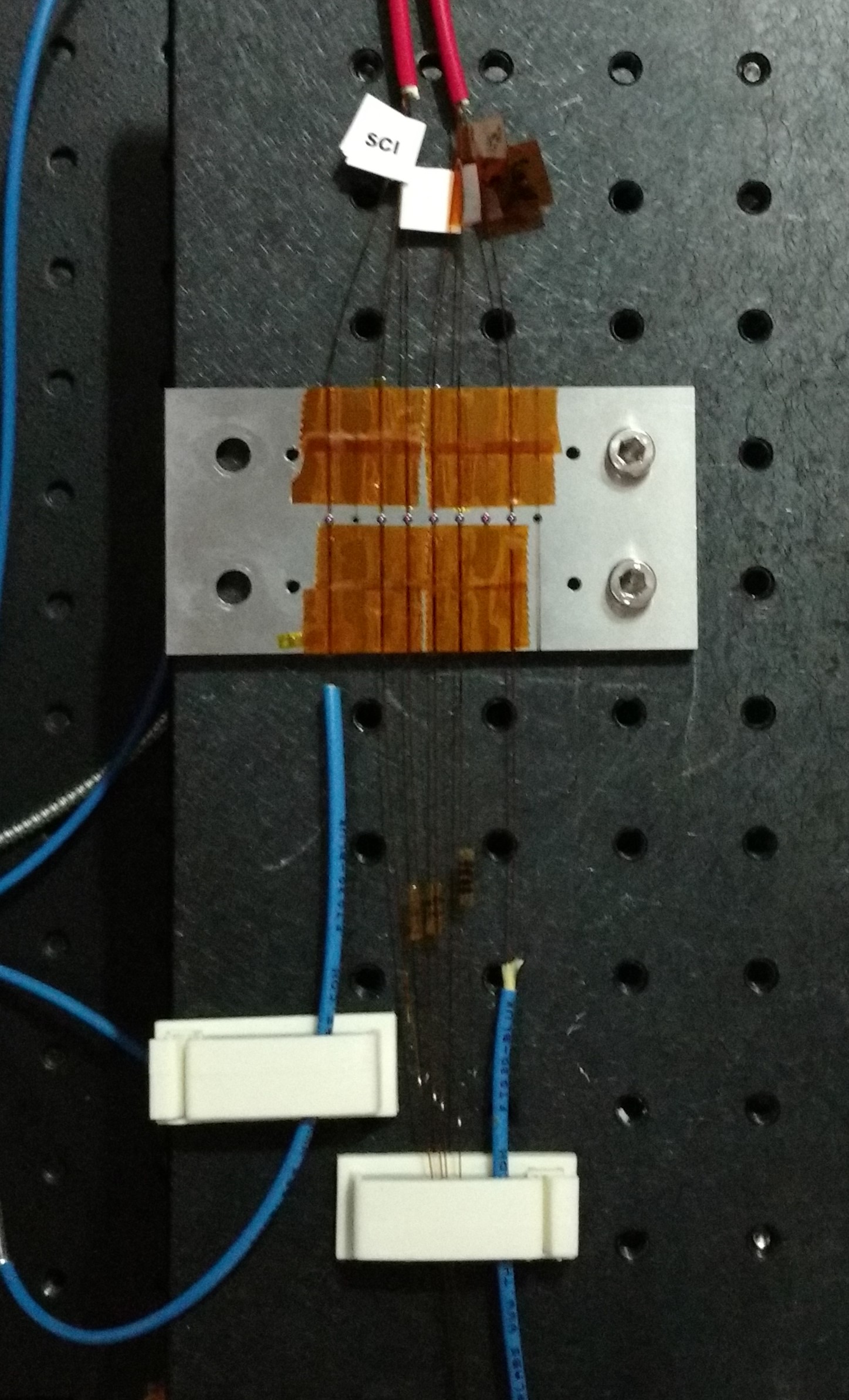}
\caption{\label{fig:ball_lens} The HPF V groove block with the ball lenses and octagonal fibers. The telescope octagonal fibers come in from the left. The two blue fibers are the two calibration fibers (HR and HE). The red jacketing on the right is the two meter patches of octagonal fibers which are then agitated.}
\end{figure}

\subsection{Fiber Agitator}

\begin{figure}[htbp]
   \captionsetup{justification=centering}
    \begin{subfigure}[t]{0.5\textwidth}
        \centering
        \includegraphics[height=2in]{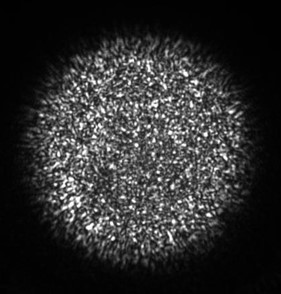}
        \caption{Fiber far field before agitation.}
    \end{subfigure}%
    ~ 
    \begin{subfigure}[t]{0.5\textwidth}
        \centering
        \includegraphics[height=2in]{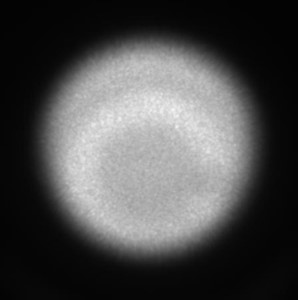}
        \caption{Fiber far field after agitation}
    \end{subfigure}
    \caption{Lab tests showing the effects of fiber agitation}\label{fig:agitation}
\end{figure}

This 2 meter patch of octagonal fiber is agitated to mitigate the modal noise. Modal noise here refers to the speckle pattern produced due to the finite number of modes traversing the fiber \cite{baudrand_modal_2001, mahadevan_suppression_2014}. This noise source is time dependent and is reduced by physically agitating the fiber. The agitation forces the temporal redistribution of the modes, hence varying the number and phase distribution of modes being traversed through by the light \autoref{fig:agitation}. The fiber agitator for HPF has not been put into regular use yet at the time of writing of this paper.

\begin{figure}[htbp]
\centering
\includegraphics[width=0.35\textwidth]{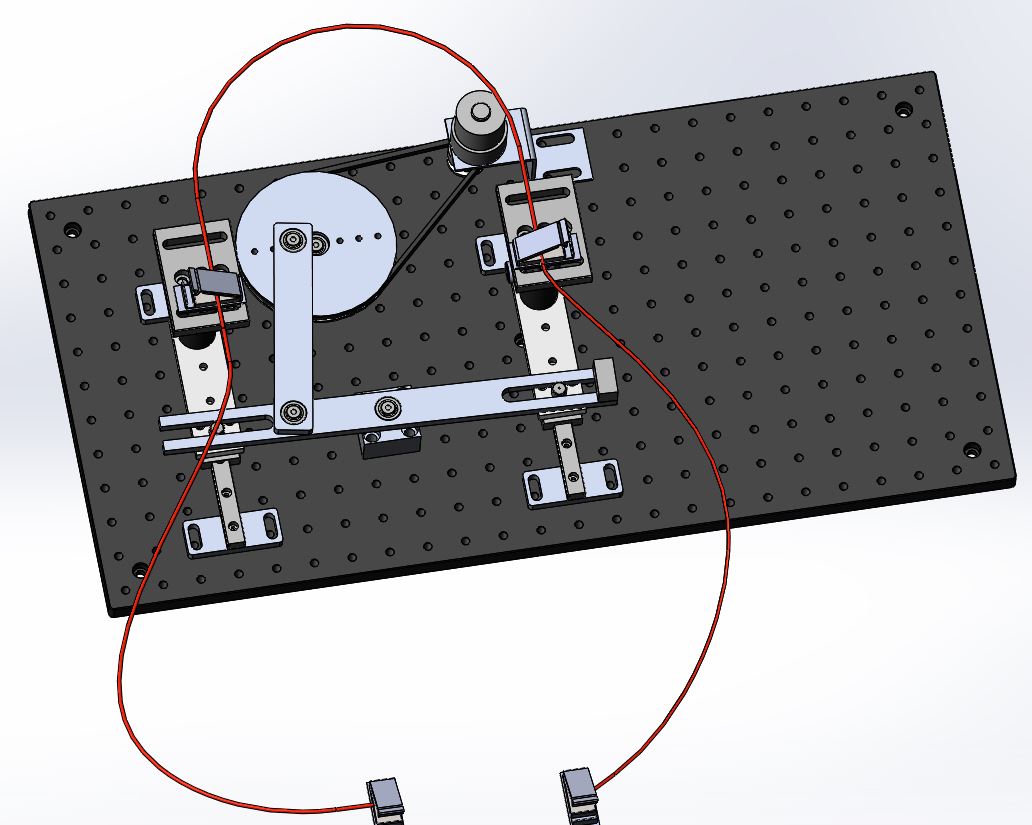}
\caption{\label{fig:agitator} The fiber agitator for HPF. The fiber here is shown in red.}
\end{figure}

\autoref{fig:agitator} shows the fiber agitator setup used for HPF. The fiber is clamped between a soft \textit{DigiKey} clamp. This holds down on the fiber without compressing or stressing it.

\subsection{Splice}
The 2 m octagonal fiber patch is then spliced on to the circular fiber from the cryostat. The splicing procedure is detailed in Section $\S$ \ref{sec:splicing}.

\subsection{Cryostat end}\label{sec:cryostat}
The fibers inside the cryostat are encased in a fiber glass sheathing. They terminate inside a fused silica puck. The procedure for preparation of this puck and the fiber inside it, along with the polishing method is detailed in Section $\S$ \ref{sec:polishing}.

The fibers pass through a \textit{NorCal} vacuum feedthrough into the vacuum cryostat. The integration of this feedthrough was done by a commercial vendor (\textit{CTech}). This is an important process because for the vacuum integrity to be maintained, the junction needs to be sealed and air tight. This is done by using epoxy to seal the junction. However, a lot of epoxy would stress the fiber and cause Focal Ratio Degradation (FRD). Therefore \textit{CTech} uses proprietary process to enclose the fibers with minimal epoxy.

\subsection{Calibration setup}
The calibration setup for HPF consists of hollow cathode lamps, a laser frequency comb\cite{osterman_near_2012} and a Fabry Perot Etalon \cite{halverson_development_2013,halverson_development_2014}. It is described in detail here \cite{halverson_habitable-zone_2014}. As shown in \autoref{fig:schematic}, an octagonal fiber goes from the cal bench to the double scrambler on the V groove block and follows the same sequence as the other octagonal fibers. There is another fiber which goes from the cal bench on to the Field Calibration Unit (FCU) \cite{lee_facility_2012}  of the  HET.

\begin{figure}[ht]
    \centering
    \begin{subfigure}[b]{0.5\textwidth}
        \centering
        \includegraphics[height=2in]{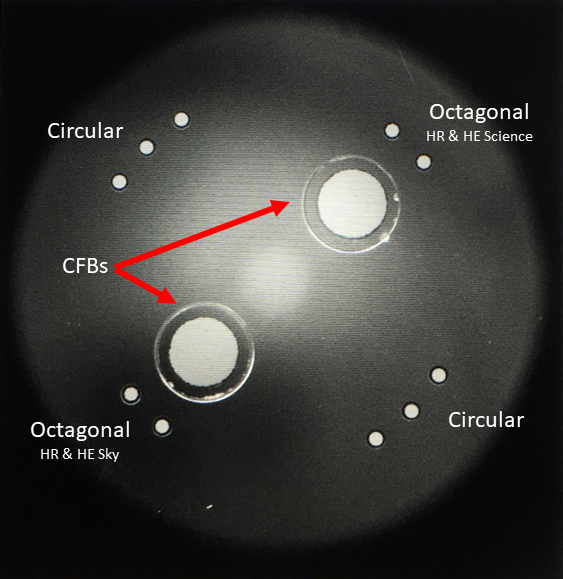}
        \caption{Image of the HPF fiberhead}
    \end{subfigure}%
    ~ 
    \begin{subfigure}[b]{0.5\textwidth}
        \centering
        \includegraphics[height=1.5in]{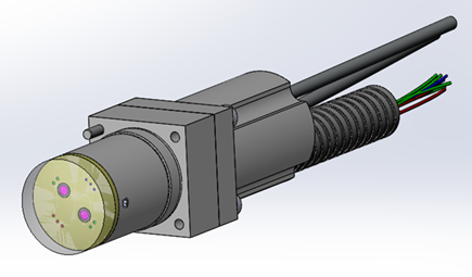}
        \caption{\textit{SolidWorks} model of the fiberhead}
    \end{subfigure}
    \caption{The fiberhead for HPF showing the HR and HE science and sky octagonal fibers. The circular fibers are spares. Currently only the HR mode is operational. }\label{fig:fiberhead}
\end{figure}

\section{HPF Fiberhead at HET Prime Focus}\label{sec:telescope_end}

\subsection{Fiberhead}

 The telescope focuses the light on to the HPF fiber head. The fiber head shown in \autoref{fig:fiberhead} has the 2 coherent fiber bundles (CFBs), 4 Octagonal fibers (Science + Sky for each mode), and 6 spare Circular fibers. This was manufactured by \textit{Berlin Fibers}, and is made out of stainless steel and modeled after the VIRUS IFU heads. The fiberhead is mounted in the Integrated Head Mounting Plate (IHMP) as part of the Prime Focus Instrument Package (PFIP)\cite{vattiat_design_2014}. HET has a focal length of about 36500 mm, and hence a plate scale of about 5.65 $^{\prime\prime}$/mm. The HPF fibers are of 300 $\mu$m or about 1.7 $^{\prime\prime}$ on sky.

\begin{figure}[hb]
\centering
\includegraphics[width=0.35\textwidth]{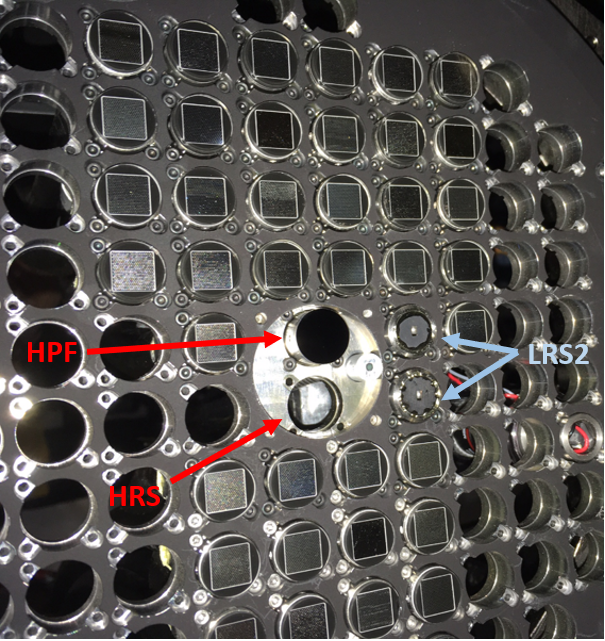}
\caption{\label{fig:ihmp} The IHMP at HET showing the HPF fiberhead and slot for HRS fiberhead at the prime focus, along with the IFUs for LRS2 and slots for VIRUS.}
\end{figure}

The IHMP at HET consists of the input fiber heads for HPF and High Resolution Spectrograph (HRS) at prime focus. A shown in \autoref{fig:ihmp}, there is the Low Resolution Spectrograph 2 (LRS2) blue and red \cite{lee_lrs2:_2010} Integral Field Unit (IFU) below that. The off axis plane consists of slots for the Visible Integral-field Replicable Unit Spectrograph (VIRUS) IFUs \cite{hill_virus:_2012}. The HPF bandpass is from 810 -- 1280 nm, hence we wanted to implement a filter for HPF, to block visible light since that would increase the scattered light inside the cryostat. The optimum location for this would be before the light enters the fiber. However, VIRUS which forms majority of the focal plan operates in the visible. Therefore a reflective filter would cause a huge scattered light problem for the other instruments. To deal with this, we installed a lens made of RG695 glass from \textit{Schott}, i.e. an absorptive glass. This will be further discussed in Section $\S$ \ref{sec:rg695}. 

\subsection{RG695 lens}\label{sec:rg695}
\begin{figure}[H]
\centering
\includegraphics[width=0.5\textwidth]{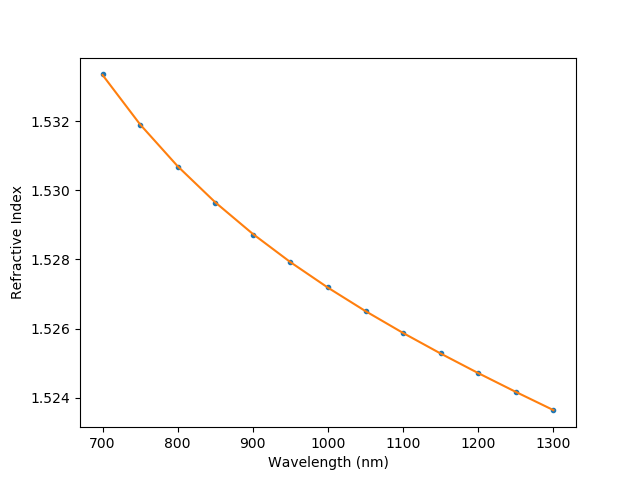}
\caption{\label{fig:rg695_refractive} The measured refractive index of the material showing the data and the Sellmeier fit.}
\end{figure}

The lens was manufactured from Schott RG695 glass by \textit{BMV Optics}. This is a long-pass glass and blocks all ($>$OD 5.0) light short of 695 nm. The lens was also AR coated to reduce Fresnel reflections. The precise refractive index of this material was not available for the HPF bandpass, and hence we first had the refractive index measured across our bandpass by \textit{M$^3$ Measurement Solutions}. Their results are shown in \autoref{fig:rg695_refractive} and \autoref{tab:refractive index}. The final lens was manufactured from the same batch of glass.The fit using Sellmeier coefficients is also shown. The residues are of the order 10$^{-5}$ -- 10$^{-6}$. The Sellmeier coefficients for the formula shown below are in \autoref{tab:sell_coeff}.

\begin{table}[!htb]
    \begin{minipage}{.5\linewidth}
    \centering
    \caption{Sellmeier Coefficients for RG695}
    \label{tab:sell_coeff}
    \begin{tabular}{|l|l|}
    \noalign{\hrule height 2pt}
    Coefficient & Value       \\ \hline
    A           & 1.0628496   \\
    B1          & 0.70484958  \\
    C1          & 0.010106028 \\
    B2          & 0.55751504  \\
    C2          & 0.012509802 \\
    B3          & 638.66496   \\
    C3          & 88905.599  \\
    \noalign{\hrule height 2pt}
    \end{tabular}
    \end{minipage}%
    \begin{minipage}{.5\linewidth}
      \centering
      \caption{Refractive Index of RG695 }
      \label{tab:refractive index}
      \begin{tabular}{|l|l|l|}
      \noalign{\hrule height 2pt}
      Wavelength (nm) & Refractive Index & Temperature ($^\circ$C) \\ \hline
      700             & 1.53336          & 22.02                 \\
      750             & 1.53190          & 22.14                 \\
      800             & 1.53069          & 21.79                 \\
      850             & 1.52965          & 22.37                 \\
      900             & 1.52872          & 22.00                 \\
      950             & 1.52792          & 22.08                 \\
      1000            & 1.52719          & 22.15                 \\
      1050            & 1.52651          & 21.90                 \\
      1100            & 1.52587          & 22.39                 \\
      1150            & 1.52529          & 22.01                 \\
      1200            & 1.52472          & 22.32                 \\
      1250            & 1.52415          & 22.23                 \\
      1300            & 1.52363          & 21.92                \\
      \noalign{\hrule height 2pt}
      \end{tabular}
    \end{minipage} 
\end{table}

\begin{equation}
n^2(\lambda) = A
+ \frac{B_1 \lambda^2 }{ \lambda^2 - C_1}
+ \frac{B_2 \lambda^2 }{ \lambda^2 - C_2}
+ \frac{B_3 \lambda^2 }{ \lambda^2 - C_3}
\end{equation}

\begin{table}[]
\centering

\end{table}

\begin{figure}[t]
    \centering
    \begin{subfigure}[b]{0.5\textwidth}
        \centering
        \includegraphics[height=2.25in]{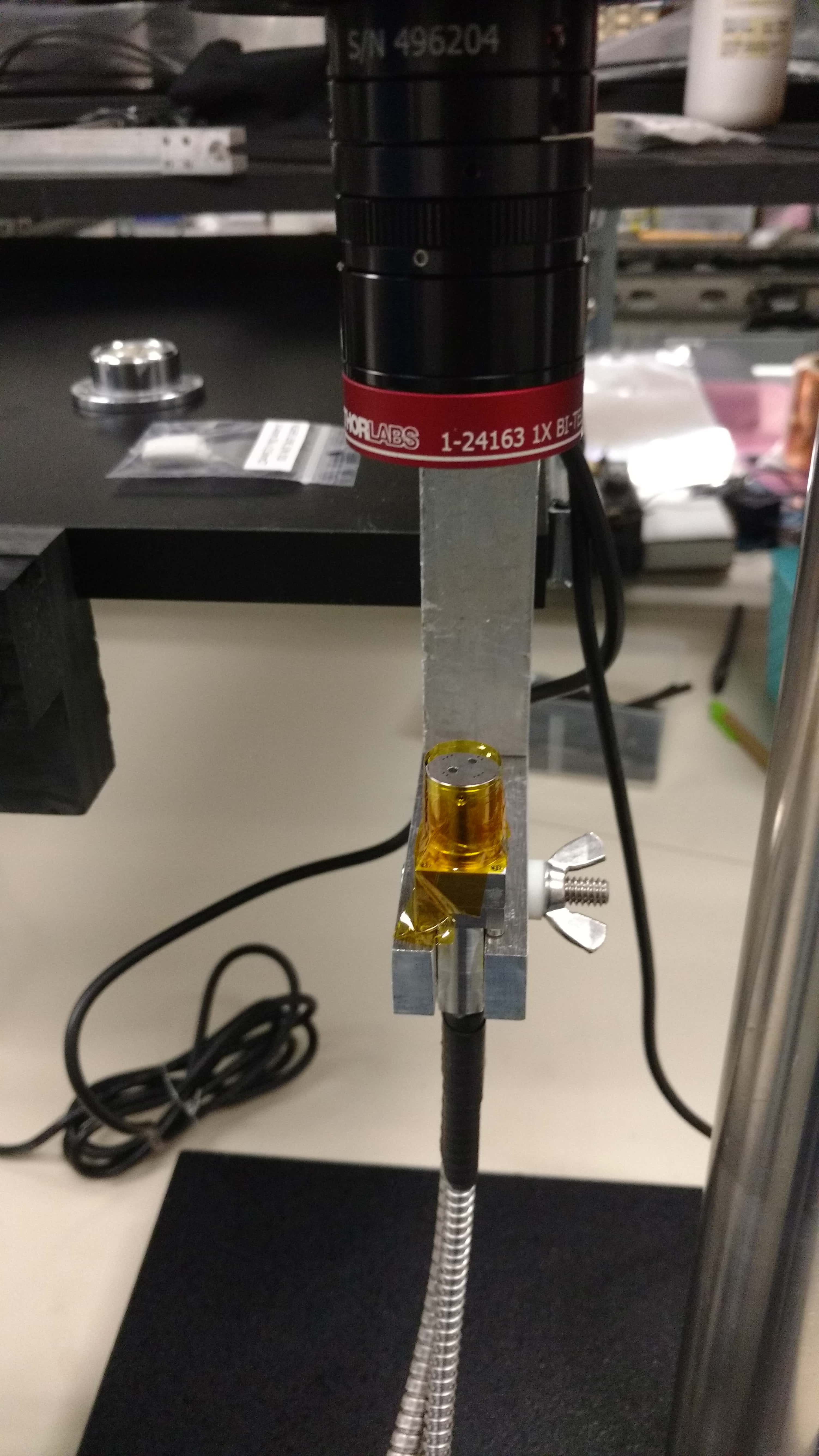}       \caption{Microscope setup to image the fiberhead during gluing.}
    \end{subfigure}%
    ~ 
    \begin{subfigure}[b]{0.5\textwidth}
        \centering
        \includegraphics[height=2.25in]{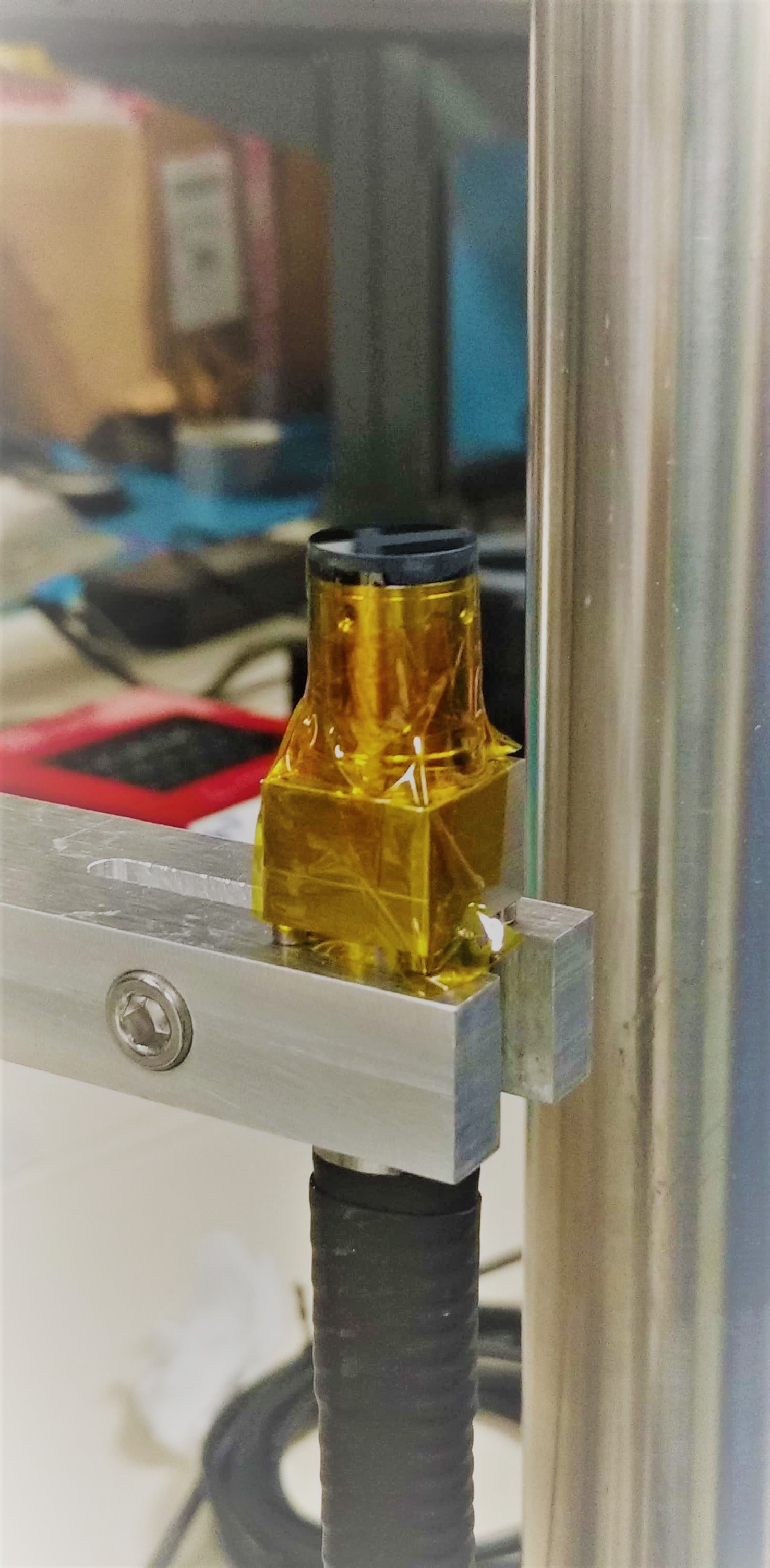}
        \caption{RG695 lens glued on top.}
    \end{subfigure}
    \caption{Gluing of the RG695 lens on the fiberhead}\label{fig:gluing_rg695}
\end{figure}

These numbers were then used in \textit{Zemax OpticStudio} to design the lens (with help from Phillip McQueen, UT to ensure the system was parfocal with both VIRUS and the HRS). The lens has a thickness of 2.5 mm and is very slightly curved to account for the beam displacement due to the lens glass itself. The lens is made to match the focal plane for the PFIP, i.e. the HPF fiberhead is parfocal with the rest of the field.The radius of curvature of the lens is about half a meter. The edge diameter of the lens matches the fiberhead dimensions to help with integration. 
To glue the lens on to the fiberhead we used \textit{Epotek 301-2} room temperature cure epoxy. The epoxy was first de-gased in a vacuum chamber to minimize the possibility of bubbles on any of the fiber or CFB inputs due to trapped air between the glass and the metal surface. We first covered the fiberhead sides using Kapton tape to ensure that none of the epoxy slid down the side of the fiberhead. This was required for a snug fit for the fiberhead in the IHMP slot. We installed our point source microscope (PSM) with a \textit{Thorlabs} telecentric lens on top of the fiberhead to document the process and verify concentricity. The telecentric lens gives a large field of view. \autoref{fig:gluing_rg695} shows the setup. Post gluing the lens we used a Teflon weight to hold down the lens as the epoxy was left to dry over 2 days.

\subsection{CFB and HPF Acquisition Camera}
\begin{figure}[hb]
\centering
\includegraphics[width=0.3\textwidth]{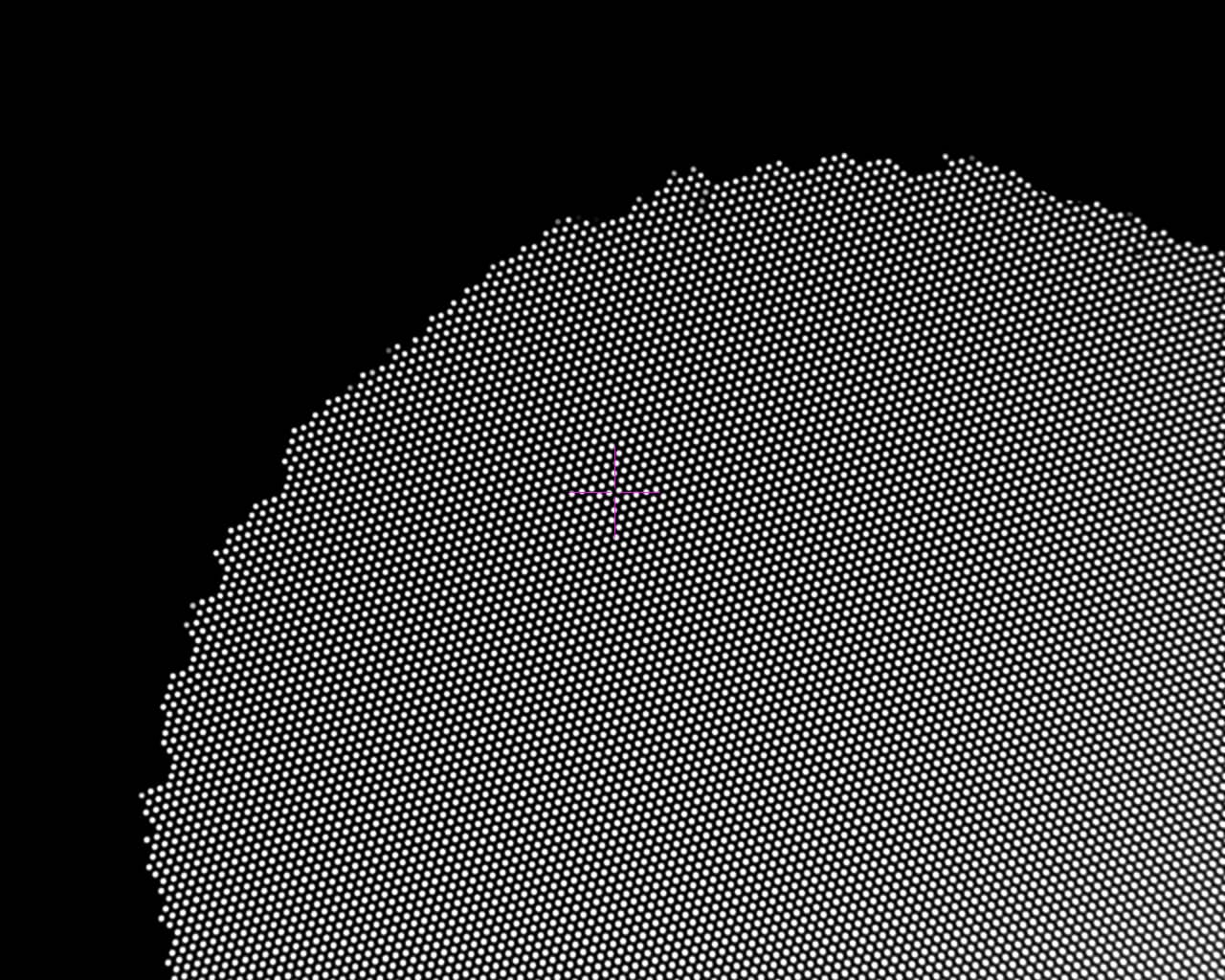}
\caption{\label{fig:cfb} An image of part of the CFB using a point source microscope when back illuminated.}
\end{figure}

The CFBs are used for accurate pointing of the telescope. Manufactured by \textit{Schott}, each CFB consists of about 18000 fibers of 12 $\mu$m diameter each. This functions as a 18 kpx image on the sky (\autoref{fig:cfb}). These provide for a means to measure the exact position of the star on the fiberhead, which then helps to move the stellar PSF from the CFB fibers to the actual science octagonal fiber. The on sky size of each of these fibers is about 70 mas. Considering typical seeing conditions, the stellar PSF extends across multiple fibers which allows us to centroid the PSF to better than the size of individual fibers. To do this centroiding, we designed a separate imaging system which can image the CFB at high cadence - HPF Acquisition Camera (HAC).

\begin{figure}[]
    \centering
    \begin{subfigure}[b]{0.8\textwidth}
        \centering
        \includegraphics[width=0.85\textwidth]{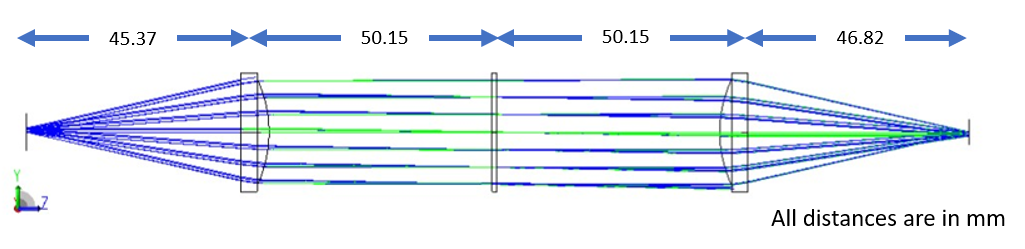}       \caption{Optical design of the HPF Acquisition Camera.}
    \end{subfigure}%
    \hfill 
    \begin{subfigure}[b]{0.8\textwidth}
        \centering
        \includegraphics[width=0.85\textwidth]{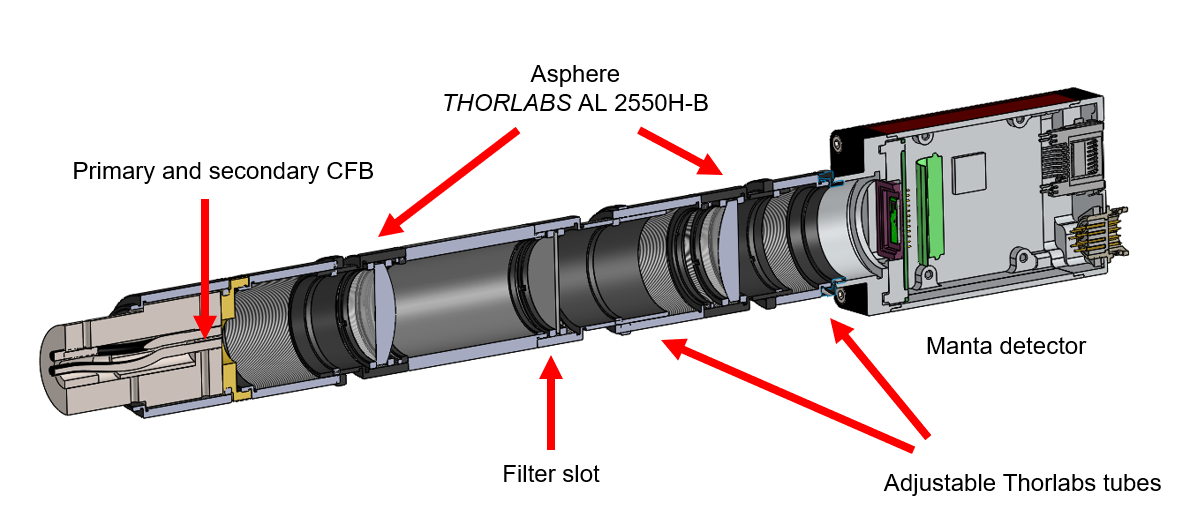}
        \caption{Opto - mechanical setup for HPF Acquisition Camera.}
    \end{subfigure}
    \caption{The classic 4f imaging system used for the HPF Acquisition Camera. It includes a filter slot at the stop, which currently contains an 857 nm filter.}\label{fig:nirbib}
\end{figure}

The HPF Acquisition Camera is a classic 4f imaging system. The system design is shown in \autoref{fig:nirbib}. We assembled this imaging system using \textit{Thorlabs} lens tubes and aspheres AR coated in the HPF bandpass. The detector used is a \textit{AVT Manta} G-223N NIR detector. The fiberhead has two CFBs, and the CFB closest to the science fiber is on axis, whereas the other one is separated by 2.96 mm (center to center). The acquisition camera is intended for commissioning and early operations-to be replaced by a larger field of view camera down the line.

\begin{figure}[htbp]
\centering
\includegraphics[width=0.6\textwidth]{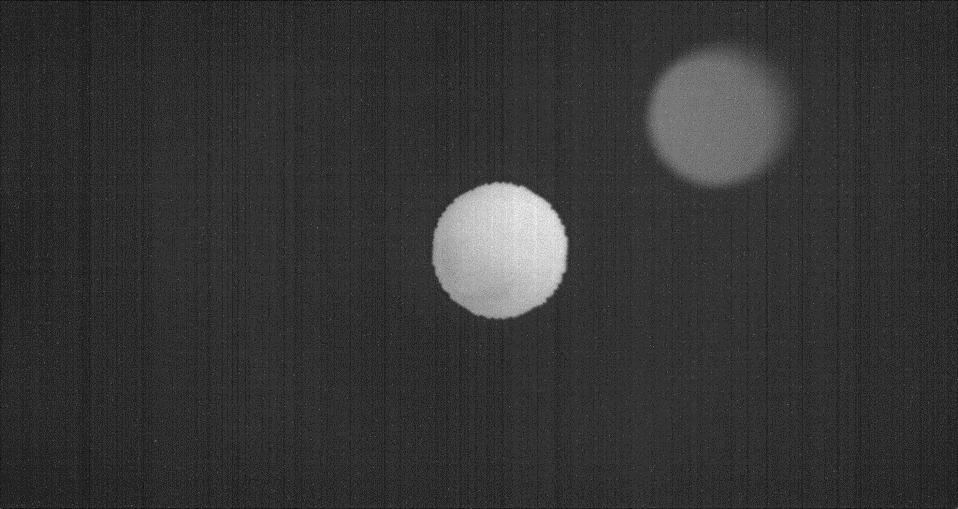}
\caption{\label{fig:nirbib_image} The image on the Manta camera showing both the CFBs. Consistent with the optical simulations, the off axis CFB shows significant aberration. However this off axis CFB aberration was deemed acceptable for design simplicity.}
\end{figure}

To mount the lenses we decided to use off the shelf 1 inch \textit{Thorlabs} SM1 lens tubes which can be adjusted to compensate for focus. This made the installation easier. To improve the PSF of the on axis fiber by mitigating chromatic aberration, we also use an 857 nm \textit{Semrock} filter. This filter can be swapped out for a different one if need be, and the relatively narrow bandpass ensures that even bright stars can be acquired and guided with the 10m HET. While designing this system we optimized to minimize the distortion for the primary (on axis) CFB, and also its telecentricity. The distortion and telecentricity measure from \textit{Zemax} for the primary CFB are 0.07 \% and 0.3 degrees respectively.  The image performance is shown in \autoref{fig:nirbib_image}.

\section{Fiber preparation}

\begin{figure}[t]
    \centering
    \begin{subfigure}[b]{0.5\textwidth}
        \centering
		\includegraphics[width=0.7\textwidth]{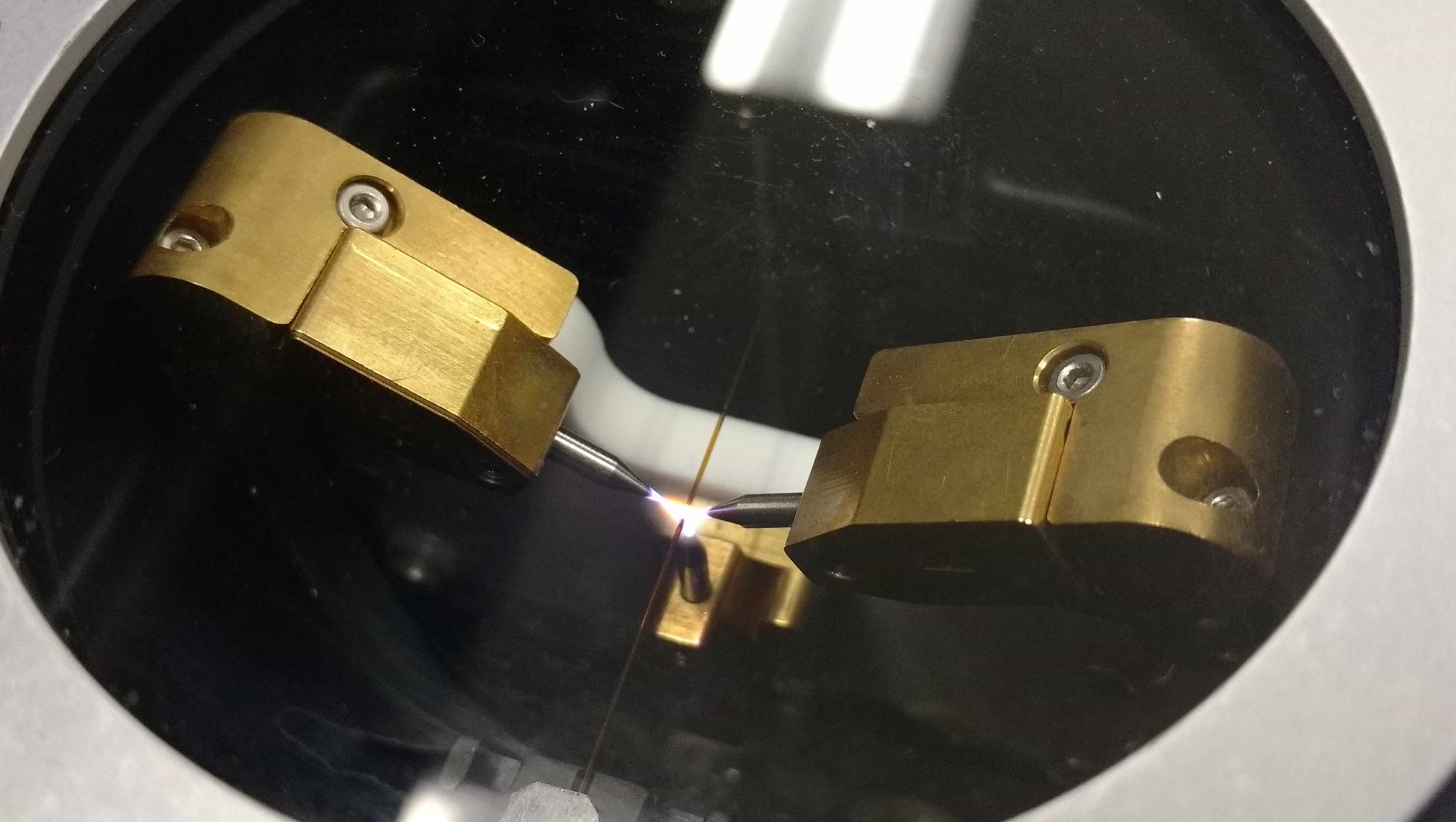}
		\caption{\label{fig:stripper} Plasma arc stripping the HPF fiber}
    \end{subfigure}%
    ~ 
    \begin{subfigure}[b]{0.5\textwidth}
        \centering
\includegraphics[width=0.7\textwidth]{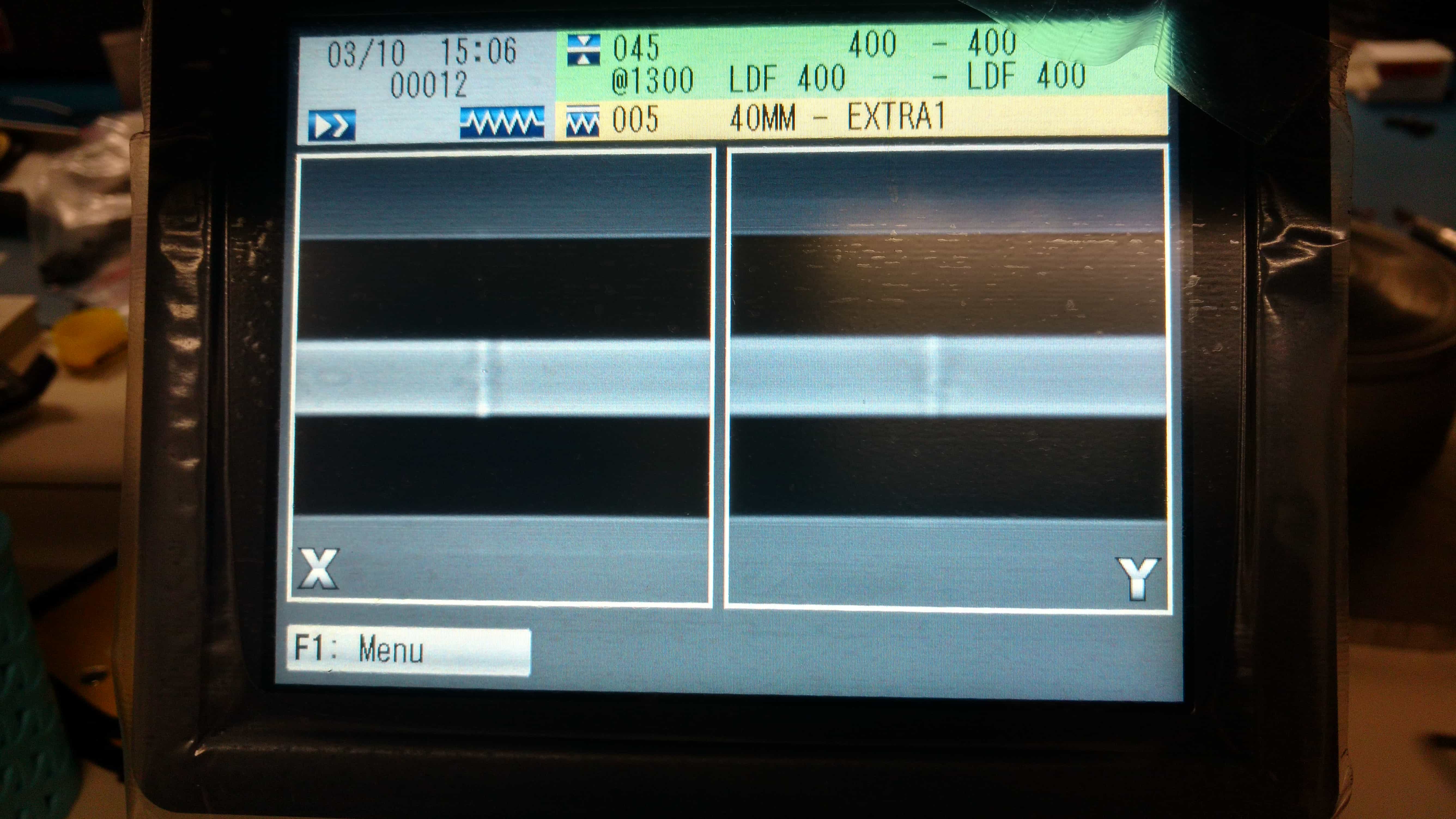}
\caption{\label{fig:splicer} Splicing image during testing of the process}
    \end{subfigure}
    \caption{Preparation and splicing of the circular fiber to octagonal fiber}\label{fig:strip_splice}
\end{figure}

As discussed in $\S$ \ref{sec:cryostat}, the octagonal fiber after the fiber agitator is spliced on to the circular cryostat fiber.

\subsection{Splicing}\label{sec:splicing}
 This spliced circular fiber is then fed into the instrument through NorCal vacuum feedthroughs.  Apart from the scrambling gain, another reason for using the spliced section of fiber for the agitation, is that since the agitator involves a moving part it has the potential for breakage or damage. It is easier to replace the spliced section of fiber rather than the cryostat or telescope fibers.

To prepare the fibers for splicing, we first strip the polyimide buffer using plasma stripping with the \textit{3SAE FPUII} (\autoref{fig:stripper}). This clamps down the fiber in a holder and then generates a plasma arc between three electrodes. This step needed to be optimized to find the best heat and cycle settings to strip off the buffer. This was followed by cleaving the stripped fiber to get it ready for splicing. For this we used the \textit{3SAE LCC} (Liquid Clamp Cleaver). Since the HPF fibers were bigger than those generally used in these units, we needed to try different recipes to obtain a good smooth cleave with a low cut angle. The cleaver uses a liquid metal clamp to hold down the fiber at one end, and then it applies tension before cutting the fiber using a diamond blade and a backstop. 

\begin{figure}[H]
\centering
\includegraphics[width=0.3\textwidth]{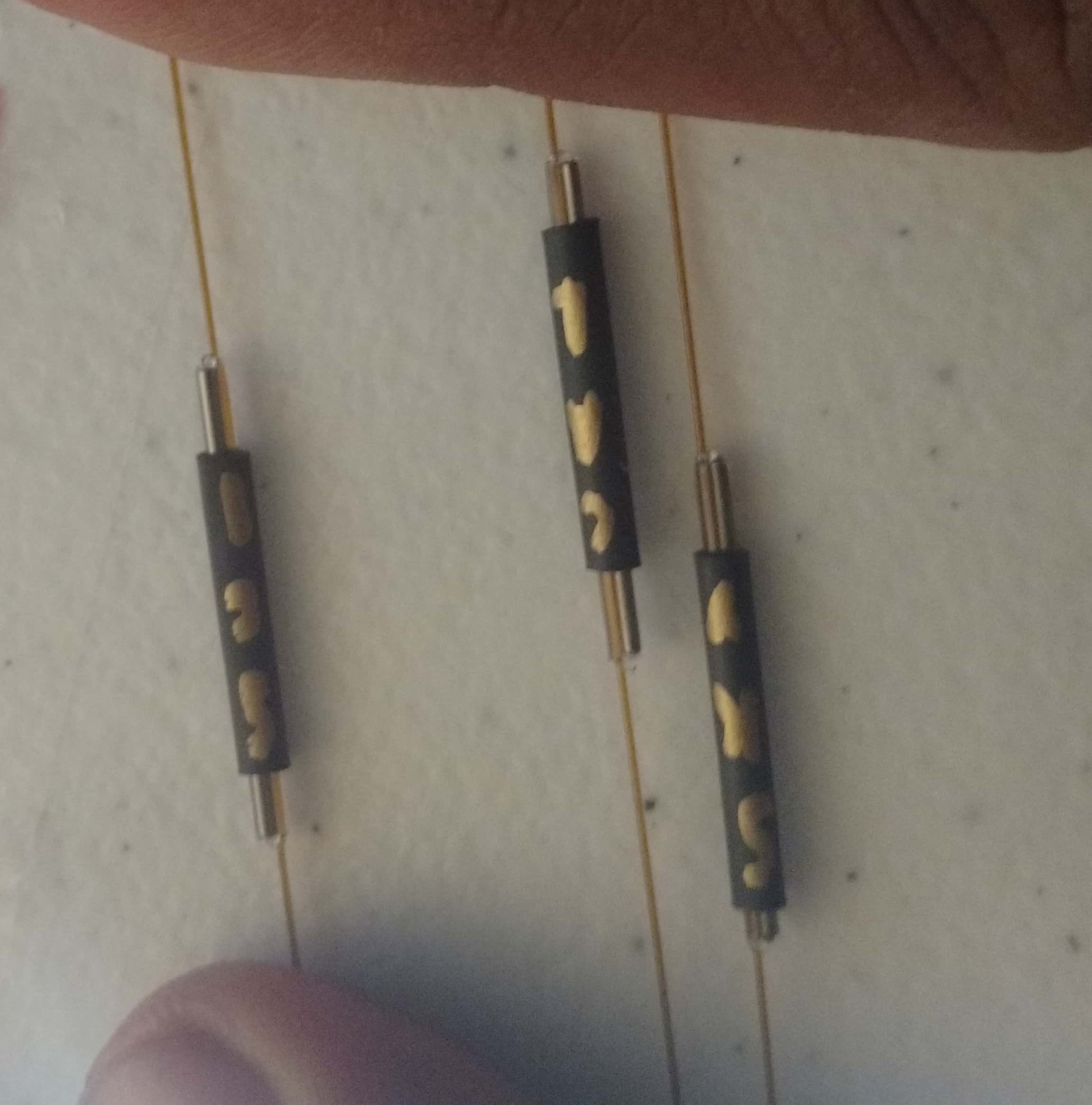}
\caption{\label{fig:splices} The three splices for HPF HR mode. }
\end{figure}

To splice the fiber we used the \textit{Fitel S183 fiber splicer}. This clamps down the fiber using fiber holders and uses a plasma arc to melt the two fibers. It then pushes them against each other to fuse them. This process has about 100 parameters that can be adjusted to ensure an optimum, minimum loss splice.  
Wee explored two different options to protect the fiber after splicing. One was re-coating the spliced section using UV cured epoxy, the other was to use a heat shrink sleeve. This is what we finally ended up using, and is shown in \autoref{fig:splices}. The splice joints are protected by a heat shrink splice sleeve from \textit{Splice Technologies}. These are then covered by a black heat shrink sleeve to prevent light leaks into the fiber. 

\subsection{Puck preparation}\label{sec:polishing}

\begin{figure}[htbp]
\centering
\includegraphics[width=0.31\textwidth]{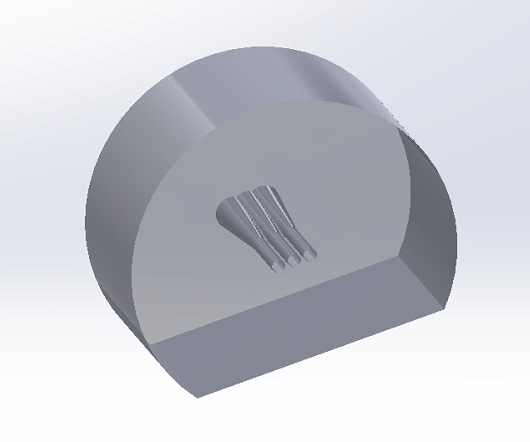}
\caption{\label{fig:silicapuck} The fused silica puck for the fiber input to the cryostat. The fused silica puck is used to terminate the three fibers. It is 10 mm in diameter and 2.5 mm in thickness. The puck also had marks etched on to it at the center line to help with positioning of the slit. These are not shown in this figure. }
\end{figure}

       The cryostat input puck for HPF was manufactured by \textit{Femtoprint}. They use femtosecond pulsed laser to change the internal structure of the fused silica, following which Hydrogen Fluoride is used to etch it. A similar Fused Silica input head is used in Maroon X\cite{seifahrt_development_2016}.
Fused Silica was used for the puck since the fiber core is also fused silica. The matching coefficient of thermal expansion reduces stresses from compression of the puck as the instrument cools down from room temperature to its operating temperature of  $\sim$ 180 K. The puck has three boreholes for the three HPF fibers, and is shown in \autoref{fig:silicapuck}. The boreholes lead up to conical openings which makes it easier to guide the fiber into the holes. The D shape is to ensure and fix the clocking of the puck in the input optics tube.

\begin{figure}[b]
    \centering
    \begin{subfigure}[b]{0.5\textwidth}
        \centering
	\includegraphics[width=0.4\textwidth]{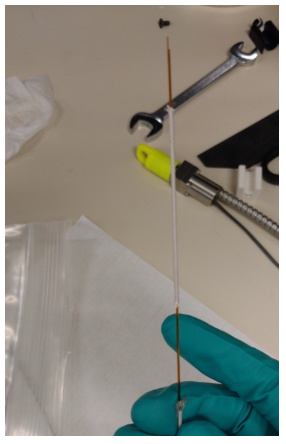}
	\caption{\label{fig:jacketing} Fiber glass jacketing used to protect fiber}
    \end{subfigure}%
    ~ 
    \begin{subfigure}[b]{0.5\textwidth}
        \centering
\includegraphics[width=0.7\textwidth]{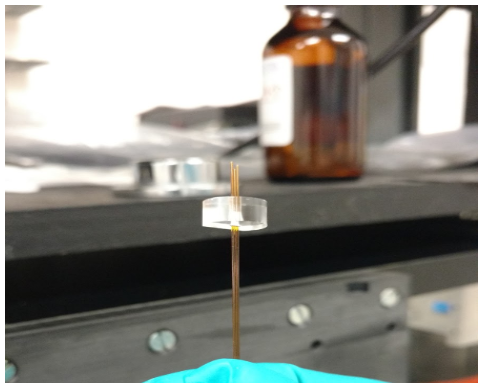}
\caption{\label{fig:puck_fiber} Inserting the fibers into the puck}
    \end{subfigure}
    \caption{Preparation of puck for epoxiing}
\end{figure}

The steps followed to prepare the puck are :- 
\begin{enumerate}
\item Use pressurized air or gauge pin to clear boreholes. It is important to this to ensure that the channels are clear. It is very difficult to extricate the fiber if it breaks or gets stuck inside the puck. Note that though gauge pins are useful, they can chip the output face of the puck, and hence must be used sparingly.

\item Hold the flat of the D parallel to the bench and insert three fibers, one after the other. Have the fibers protruding such that the jacketing is in contact with the puck. This makes it easier to hold on to the fiber - puck. If there is too much bare fiber, can add additional jacketing to protect fiber (\autoref{fig:jacketing}).

\item To ensure that the D puck fits in its mount, we need to make sure that none of the epoxy drips down its sides while it dries. Therefore cover it with \textit{Kapton} tape to protect the sides (\autoref{fig:puck_tape}). Can remove the tape after the epoxy dries. 

\item Mount the puck on to the custom holder and clamp down. Ensure that the fibers are not caught in the clamp.

\item Decrystallize \textit{Epotek 301-2} epoxy by warming it up in lukewarm water. Prepare epoxy by recommended weight ratio.

\item De-gas epoxy in a vacuum chamber to remove air bubbles that formed during the mixing of the epoxy (\autoref{fig:degas}).

\begin{figure}[h]
    \centering
    \begin{subfigure}[b]{0.5\textwidth}
        \centering
	\includegraphics[width=0.8\textwidth]{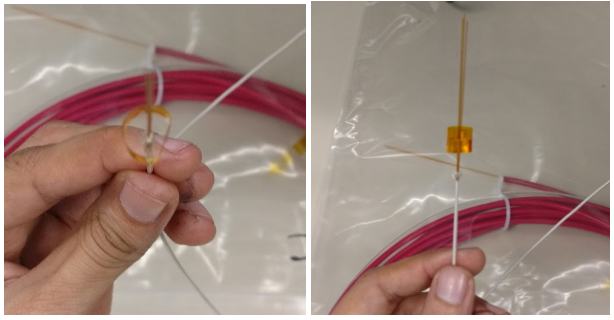}
	\caption{\label{fig:puck_tape} Kapton tape to prevent epoxy overflow}
    \end{subfigure}%
    ~ 
    \begin{subfigure}[b]{0.5\textwidth}
        \centering
\includegraphics[width=0.8\textwidth]{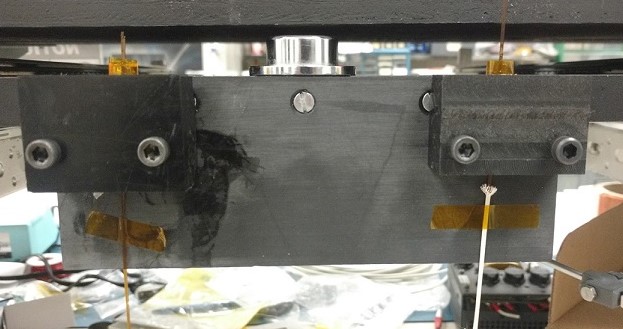}
\caption{\label{fig:puck_stand} Clamping puck in custom made stand}
    \end{subfigure}
    \caption{Preparation of puck for epoxiing}
\end{figure}

\item Use syringe to load up epoxy. Be careful to do so without introducing too many air bubbles. Apply epoxy by inserting down the protruding fibers.

\item Once applied, slide the fiber within the puck to ensure that the epoxy percolates down in to the puck bore. 

\item Tape fiber on to the clamp to hold it there, as shown in \autoref{fig:puck_stand}.

\begin{figure}[htbp]
\centering
\includegraphics[width=0.3\textwidth]{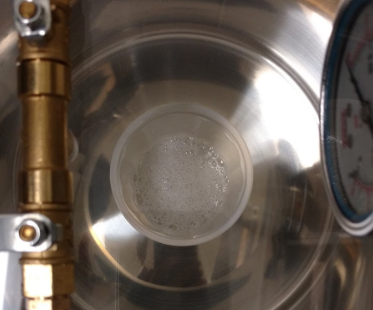}
\caption{\label{fig:degas} De-gasing the epoxy in a vacuum chamber}
\end{figure}

\item Check with bright light source for any light leaks or breaks in fiber due to micro cracks. If so, cleave off fiber a few cm before the crack and restart the process.

\item This epoxy takes about 2 days to cure. Therefore after 2 days we carefully removed the clamp and the Kapton tape and now have the puck ready for polishing (\autoref{fig:puck_epoxy}). 

\begin{figure}[H]
\centering
\includegraphics[width=0.2\textwidth]{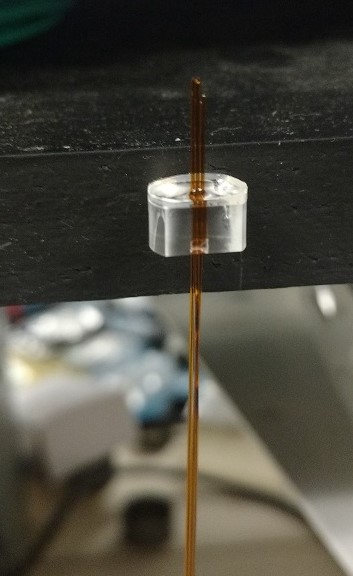}
\caption{\label{fig:puck_epoxy} The layer of epoxy on the puck after removing the Kapton tape.}
\end{figure}
\end{enumerate}

\begin{figure}[b]
    \centering
    \begin{subfigure}[b]{0.5\textwidth}
        \centering
	\includegraphics[width=0.7\textwidth]{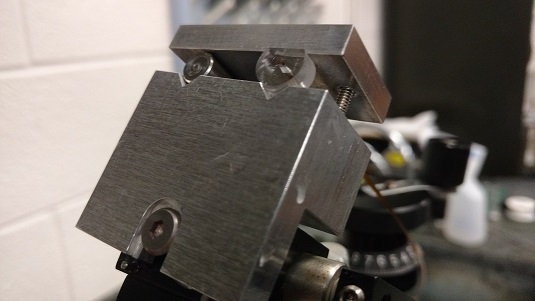}
	\caption{\label{fig:polishing_jig} Aluminium polishing jig holding the fiber puck and a steel cylinder for counter clamp.}
    \end{subfigure}%
    ~ 
    \begin{subfigure}[b]{0.5\textwidth}
        \centering
\includegraphics[width=0.7\textwidth]{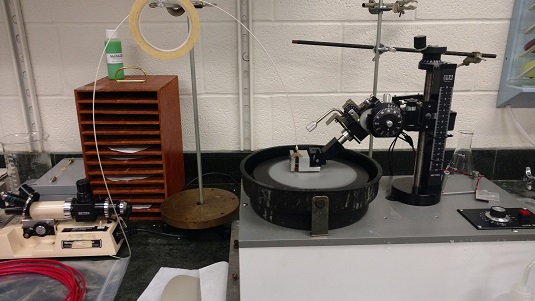}
\caption{\label{fig:polishing_wheel} Supporting the fiber with lab stand. Need to make sure that the fiber is not bent more than its natural bend radius.}
    \end{subfigure}
    \caption{Polishing of the fiber puck}
\end{figure}

To polish the puck we use a polishing wheel with a calibrated arm for angle and height adjustment (\autoref{fig:polishing_wheel}). To mount the puck on to this we manufactured an aluminium polishing jig (\autoref{fig:polishing_jig}). The arm has an electrical contact which is connected to a LED. Therefore when the polishing grit wears down a layer of epoxy, the contact is broken and the LED lights up, signalling that we can lower the arm to continue polishing.  

The steps followed for polishing are :- 
\begin{enumerate}
\item Mount puck in the polishing jig as is shown in \autoref{fig:polishing_jig}.
\item To support the fiber we used \textit{Apiezon Q} vacuum compatible putty.
\item Wipe down polishing wheel with distilled water and wash the polishing film with it as well.
\item Start with coarse grain ($\sim$ 60 $\mu$m) to grind down the protruding fiber till contact is made with the epoxy layer. Squirt distilled water to wash away the ablated material that is being ground down. 
\item If there is too thick a layer of epoxy, continue using 60 $\mu$m.
\item Then smoothen the surface by using a 12 $\mu$m film. 
\item After this use 5 or 3 $\mu$m for finer removal, post which can use 1 $\mu$m once the epoxy layer starts thinning down.
\item Can apply finishing touches with 0.5 or 0.3 $\mu$m. We used the Ultrapol film from \textit{Ultra Tec} after this for our final polish.
\item Be careful to wash the films before and after use with distilled water. It is important to this since even a tiny speck of dust can ablate through the puck and cause scratches or gashes.
\item After every couple or so films, inspect the puck using a fiber inspection microscope. Before doing so, can wipe the surface with a lint free wipe and ethanol, and can blow away the excess water using clean pressurised air. 
\item Every few rounds or so, also check for light leaks and cracks using bright light source. It is important to do so since any jerk or slippage during the polishing can break or potentially crack the fiber.
\end{enumerate}

The various kinds of polishing films and compounds we tried were - Silicon Carbide, Aluminium Oxide, Diamond, Cerium Oxide powder (\textit{Rhodox 76}), and Alumina. We found that a combination of Silicon Carbide for the coarse grits, and Aluminium Oxide and diamond for the finer ones worked best for our puck, fiber and epoxy combination.

\section{Inspection mechanism}
To inform the overall quality of the full HPF fiber train, we performed extensive testing of all of the HPF fibers in two main areas. First, we developed an automated setup to precisely measure the Fiber-Focal-Ratio degradation of the HPF fibers. Second, we employed a Zygo Optical Profilometer to assess the surface roughness of the fiber polishing and slit epoxying. These tests are further described below.

\subsection{Fiber Focal Ratio Degradation (FRD)}
% p#3:
We give a brief overview of the FRD measurement system we developed to measure the FRD properties of the HPF fibers. We refer the reader to a more detailed discussion of this system in a forthcoming publication, Stefansson et al. 2019 (in prep).

% p#1: What is FRD?
Fiber Focal Ratio Degradation (FRD) originates from imperfections in optical fibers, including microbends and general inhomogeneities present in and around the fiber cladding/core interface. Of interest to high resolution fiber-fed spectrographs, FRD manifests as an increase in the exit angles---and decrease in output F/\#---of the output light cone \cite{ramsey_focal_1988}. This increase in exit angles can lead the output light to overfill the optics (e.g., the collimating parabola in a white-pupil echelle spectrograph design). Overfilling the optics directly leads to a loss in the overall spectrograph efficiency, and can lead to an increase in scattered light without sufficient baffling.

\subsubsection*{FRD setup}
\begin{figure}[H]
\centering
\includegraphics[width=0.85\textwidth]{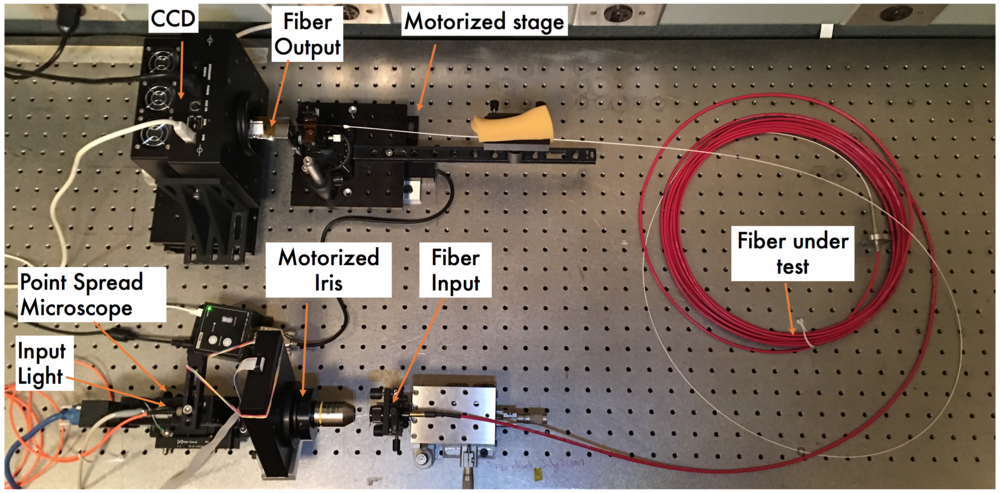}
\caption{An overview of the HPF FRD setup, which uses a Point Spread Microscope to deliver a perpendicular input light cone into the fiber under test. A motorized iris offers the ability to easily change the input F/\# at the injection end, and by mounting the output end of the fiber under test, we can precisely measure the output F/\# by measuring the distance to the CCD imager via the method of similar triangles. A more detailed further description of the setup is discussed in the text.}
\label{fig:frd_setup}
\end{figure}

In short, the cornerstone of our FRD measurement system uses a Point-Spread-Microscope (PSM) from \textit{Davidson Optronics}. \autoref{fig:frd_setup} gives an overview of the setup used to characterize the FRD properties of the HPF fibers. We use an Alphabright incoherent light source as the input light source to minimize coherent-light mode speckling at the output. We feed in the light from the Alphabright via a 50 $\mu m$ input fiber into the PSM optical train. We mount a motorized iris in the collimated space of the PSM to allow us to change the input illumination F/\# for the fiber under test. Additionally, in autocollimation mode (i.e., by removing the PSM lens objective), we can precisely collimate the PSM against the input fiber, allowing us to deliver an input light cone perpendicular to the input-end of the fiber under test. At the output end of the test-fiber, we mount the test fiber on a motorized stage to allow us to translate the output a given distance from the CCD detector. Doing so, allows us to sample slices of the output cone at a well-defined set of distances to precisely calculate the distance of the fiber tip to the detector plane, which can subsequently be converted to an output F/\# for subsequent FRD analysis. To regularize the process, we developed an automated pipeline in Python to obtain a well defined test suite of images to fully characterize the FRD of a fiber under test from $\sim$F/2.5 to $\sim$F/7. This pipeline will be further described in Stefansson et al. 2019 (in prep).

\begin{figure}[H]
\centering
\includegraphics[width=0.85\textwidth]{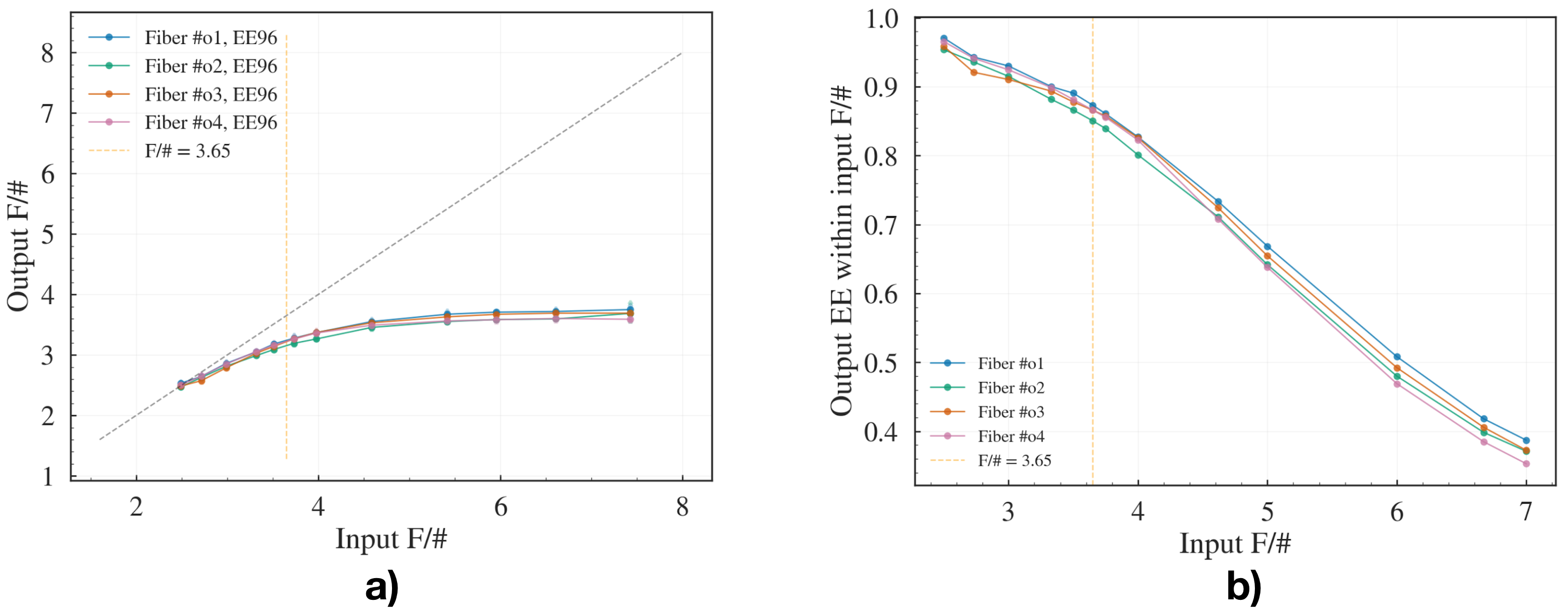}
\caption{Example FRD results of the octagonal HPF top-end fibers (40m long). a) Output F/\# as a function of input F/\#. The gray-line shows the one-to-one line if there was no FRD present. b) Encircled energy within the input f-number as measured in the output PSF as a function of input number.}
\label{fig:frd_result}
\end{figure}

 \autoref{fig:frd_result} shows an example output plot from our FRD pipeline of the final four HPF octagonal science fibers that deliver the light from the Hobby-Eberly Telescope Focal Plane to the HPF Calibration Room enclosure ($\sim$40m long fibers protected in a Stainless Steel jacketing). The left panel shows the output F/\# as a function of input F/\# for the 4 fibers as measured by our automated FRD setup. The diameter we use to calculate the output F/\# is defined as the 96\% Encircled Energy at the output PSF as measured by the CCD imager in Figure \ref{fig:frd_setup}. The right panel shows the Encircled Energy within the input F/\# in the output PSF, as a function of the injected F/\# for the same four fibers. %From the right panel, we see that at F/3.65---the F/\# of HET---the fibers all have an encircled energy of $\sim$90\%.

\subsection{Zygo Nexview Optical Profilometer}
We also utilized a \textit{Zygo} Nexview Optical Profilometer to measure the profile of the fibers after polishing, and the Nickel Slit before gluing. This is basically an interferometer and is useful to measure surface roughness and dimensions accurately. This was done to measure the slit width precisely ($\sim$ 1 $\mu$m) to pick the optimum one. The final slit we installed on HPF HR mode was 93 $\mu$m wide. We also used this to measure the v - groove blocks to ensure that the fibers on both sides of the ball lens are in the same plane. 

\begin{figure}[H]
\centering
\includegraphics[width=0.4\textwidth]{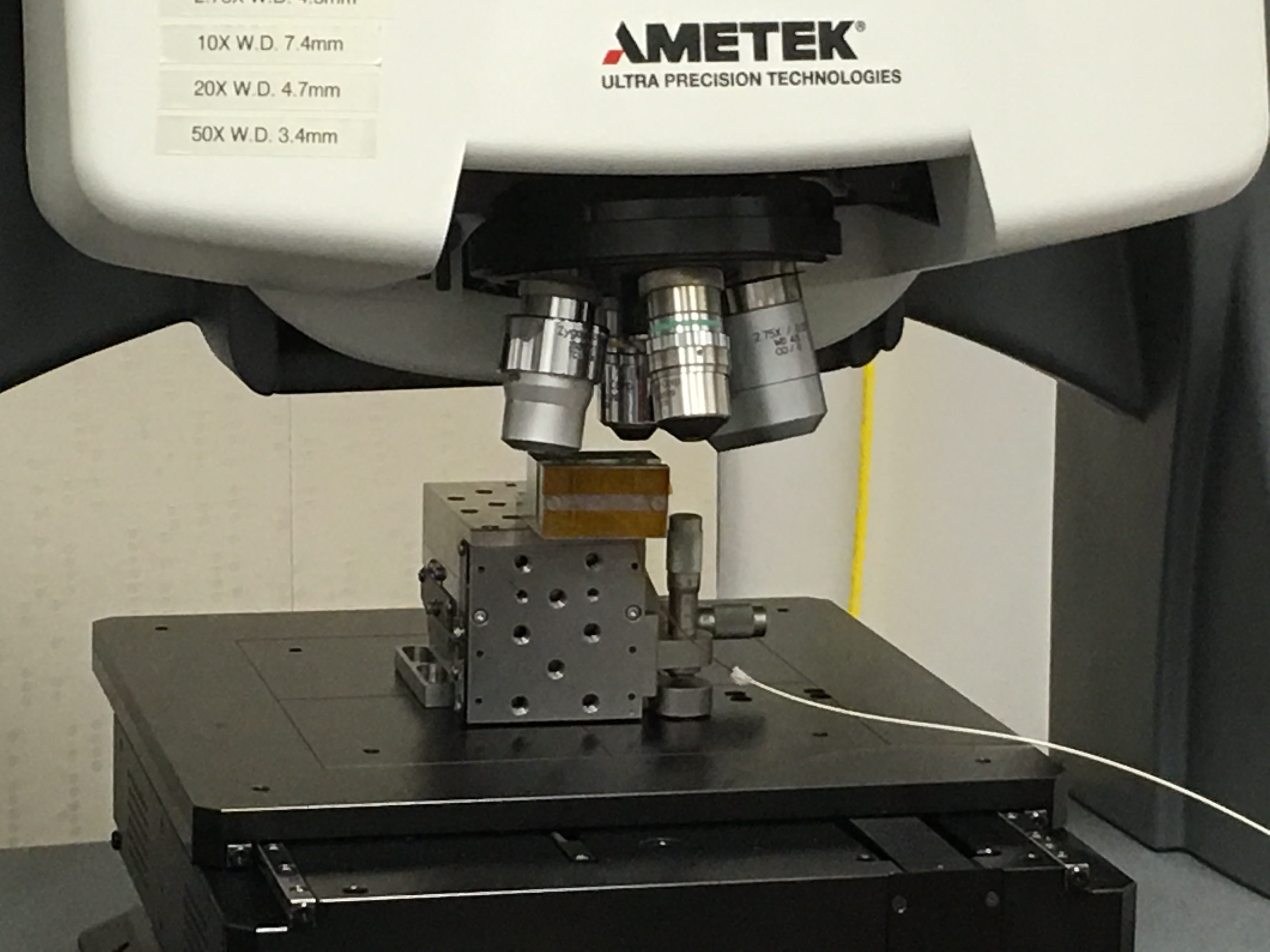}
\caption{\label{fig:zygo} Zygo interferometer measurements of the puck after polishing}
\end{figure}

\section{Slit installation and rotation}
\subsection{Installation}
\begin{figure}[H]
\centering
\includegraphics[width=0.4\textwidth]{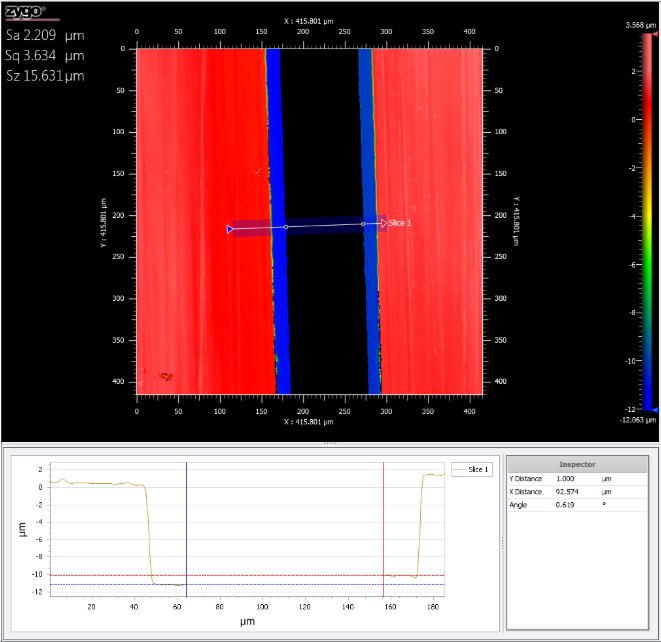}
\caption{\label{fig:slit} Nickel slit used for HPF HR mode} as measured by Zygo interferometer.
\end{figure}

We ordered 20 nickel slits of nominal width 100 $\mu$m and thickness 25 $\mu$m. These were blackened on the outside. The final slit that was used on the puck had a width of about 93 $\mu$m and is shown in \autoref{fig:slit}. To install the slit we considered using Potassium Hydroxide to bond the slit on to the puck, as explored in \cite{mahadevan_inexpensive_2008}. However this method did not work for us. We finally used \textit{Norland} 68 epoxy which requires UV curing. 

The process of installation of the slit was very difficult and required a few tries to get correct, since we did not want any of the epoxy to get on to the fiber. We used pressurized air, alcohol and cotton swabs to clean the fibers after the slit was installed.

\begin{figure}[htbp]
    \centering
    \begin{subfigure}[b]{0.5\textwidth}
        \centering
	\includegraphics[width=0.8\textwidth]{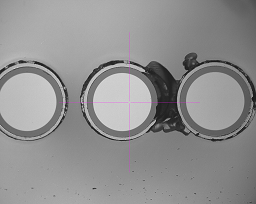}
	\caption{\label{fig:HR_fibers} The dark region is residual epoxy between the fibers}
    \end{subfigure}%
    ~ 
    \begin{subfigure}[b]{0.5\textwidth}
        \centering
\includegraphics[width=0.8\textwidth]{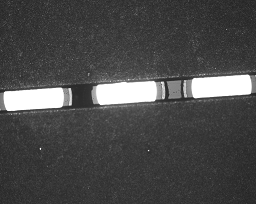}
\caption{\label{fig:HR_slit} Showing the 93 $\mu$m slit.}
    \end{subfigure}
    \caption{HPF puck before and after sluing the slit}
\end{figure}

\subsection{Rotation}
The puck had to be rotated in order to correct for the rotation of the field due to the gamma angle \cite{chaffee_astronomical_1976}. For HPF this rotation was about 4.4$^\circ$, this enabled the fiber to be flat in the cross dispersion direction.

If the slit image is not along the column pixel axis, the 2D to 1D extraction will need to interpolate across slanted pixel grid. This interpolation will reduce the radial velocity information content\footnote{RV Information content is proportional to slope, any interpolation is a convolution operation, which will reduce the slope of the spectrum.}. There will also be other artefacts in the extracted spectrum at sub-m/sec RV precision, which we would like to minimise.  A large slit rotation offset also means the sky and cal fiber will be significantly offset from the science fiber, resulting in larger PSF difference.
Even though it is not possible to remove the slit rotation across a full order, HPF's short orders enable us to reduce the median slit image rotation to be very close to zero by tilting the physical slit. 
The resultant velocity difference between the fibers after removing the slit tilt is about 40 m/s or 0.02 px. This varies by wavelength as the PSF changes.

\begin{figure}[H]
\centering
\includegraphics[width=0.4\textwidth]{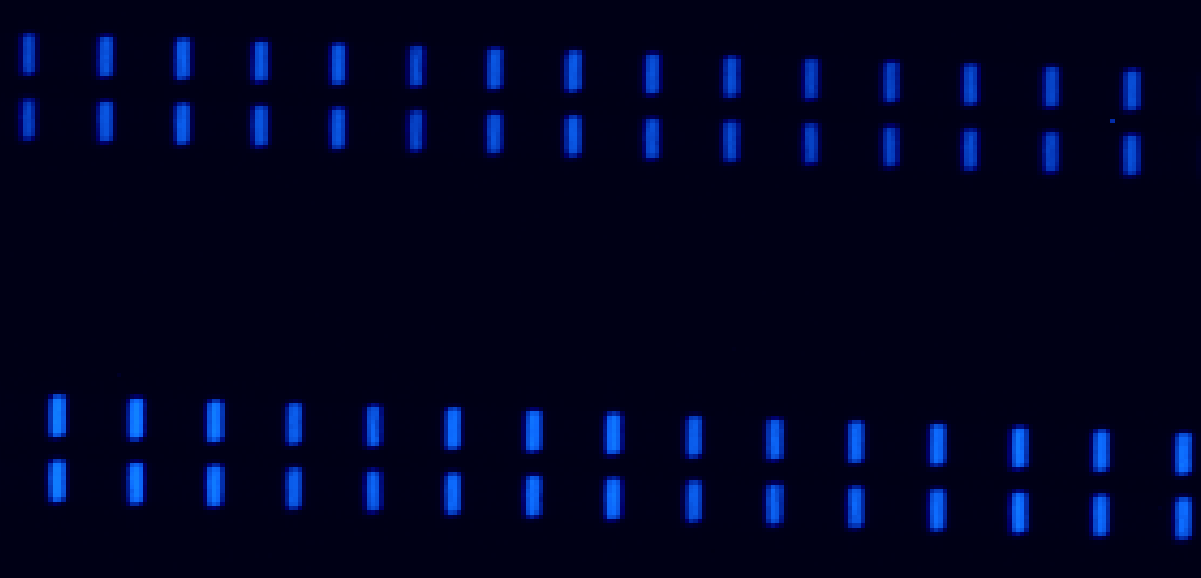}
\caption{\label{fig:rotation} Two out of three HPF fibers illuminated showing the slit alignment.}
\end{figure}

\acknowledgments % equivalent to \section*{ACKNOWLEDGMENTS}       
 
This work was partially supported by the funding from The Center for Exoplanet and Habitable Worlds. The Center for Exoplanet and Habitable Worlds is supported by The Pennsylvania State University, The Eberly College of Science, and The Pennsylvania Space Grant Consortium. We acknowledge support from NSF grants, AST1006676, AST1126413, AST1310885, and the NASA Astrobiology Institute (NNA09DA76A) in our pursuit of precision radial velocities in NIR.
 SK acknowledges support from the Braddock fellowship. GKS acknowledges support from the Leifur Eiriksson foundation.

% References
\bibliography{SPIE_fibers} % bibliography data in report.bib

\begin{thebibliography}{10}

\bibitem{k._serkowski_fabry-perot_1979}
Serkowski, K., Frecker, J.~E., Heacox, w.~D., and Roland, E.~H.,
  ``Fabry-{Perot} {Radial} {Velocity} {Spectrometer},'' in [{\em
  Instrumentation in {Astronomy} {III}}{\nolinebreak\hspace{0.1em}]},   {\bf
  0172},  130--134, International Society for Optics and Photonics (May 1979).

\bibitem{black_assessment_1980-1}
Black, D.~C. and Brunk, W.~E., ``An {Assessment} of {Ground}-{Based}
  {Techniques} for {Detecting} {Other} {Planetary} {Systems}. {Volume} 2:
  {Position} papers,''  {\bf 2124} (Feb. 1980).

\bibitem{barden_evaluation_1981}
Barden, S.~C., Ramsey, L.~W., and Truax, R.~J., ``Evaluation of some fiber
  optical waveguides for astronomical instrumentation,'' {\em Publications of
  the Astronomical Society of the Pacific}~{\bf 93},  154 (Feb. 1981).

\bibitem{heacox_application_1986}
Heacox, W.~D., ``On the application of optical-fiber image scramblers to
  astronomical spectroscopy,'' {\em The Astronomical Journal}~{\bf 92},
  219--229 (July 1986).

\bibitem{black_assessment_1980}
Black, D.~C. and Brunk, W.~E., ``An assessment of ground-based techniques for
  detecting other planetary systems. {Volume} 1: {An} overview,''  {\bf 2124}
  (Feb. 1980).

\bibitem{mayor_jupiter-mass_1995}
Mayor, M. and Queloz, D., ``A {Jupiter}-mass companion to a solar-type star,''
  {\em Nature}~{\bf 378},  355--359 (Nov. 1995).

\bibitem{baranne_elodie:_1996}
Baranne, A., Queloz, D., Mayor, M., Adrianzyk, G., Knispel, G., Kohler, D.,
  Lacroix, D., Meunier, J.-P., Rimbaud, G., and Vin, A., ``{ELODIE}: {A}
  spectrograph for accurate radial velocity measurements.,'' {\em Astronomy and
  Astrophysics Supplement Series}~{\bf 119},  373--390 (Oct. 1996).

\bibitem{mayor_setting_2003}
Mayor, M., Pepe, F., Queloz, D., Bouchy, F., Rupprecht, G., Lo~Curto, G.,
  Avila, G., Benz, W., Bertaux, J.-L., Bonfils, X., Dall, T., Dekker, H.,
  Delabre, B., Eckert, W., Fleury, M., Gilliotte, A., Gojak, D., Guzman, J.~C.,
  Kohler, D., Lizon, J.-L., Longinotti, A., Lovis, C., Megevand, D., Pasquini,
  L., Reyes, J., Sivan, J.-P., Sosnowska, D., Soto, R., Udry, S., van Kesteren,
  A., Weber, L., and Weilenmann, U., ``Setting {New} {Standards} with
  {HARPS},'' {\em The Messenger}~{\bf 114},  20--24 (Dec. 2003).

\bibitem{lovis_exoplanet_2006}
Lovis, C., Pepe, F., Bouchy, F., Lo~Curto, G., Mayor, M., Pasquini, L., Queloz,
  D., Rupprecht, G., Udry, S., and Zucker, S., ``The exoplanet hunter {HARPS}:
  unequalled accuracy and perspectives toward 1 cm s $^{\textrm{-1}}$
  precision,''  62690P (June 2006).

\bibitem{kasting_habitable_1993}
Kasting, J.~F., Whitmire, D.~P., and Reynolds, R.~T., ``Habitable {Zones}
  around {Main} {Sequence} {Stars},'' {\em Icarus}~{\bf 101},  108--128 (Jan.
  1993).

\bibitem{mahadevan_habitable-zone_2012}
Mahadevan, S., Ramsey, L., Bender, C., Terrien, R., Wright, J.~T., Halverson,
  S., Hearty, F., Nelson, M., Burton, A., Redman, S., Osterman, S., Diddams,
  S., Kasting, J., Endl, M., and Deshpande, R., ``The {Habitable}-{Zone}
  {Planet} {Finder}: {A} {Stabilized} {Fiber}-{Fed} {NIR} {Spectrograph} for
  the {Hobby}-{Eberly} {Telescope},'' {\em arXiv:1209.1686 [astro-ph]} ,
  84461S (Sept. 2012).
\newblock arXiv: 1209.1686.

\bibitem{stefansson_versatile_2016}
Stefansson, G., Hearty, F., Robertson, P., Mahadevan, S., Anderson, T., Levi,
  E., Bender, C., Nelson, M., Monson, A., Blank, B., Halverson, S., Henderson,
  C., Ramsey, L., Roy, A., Schwab, C., and Terrien, R., ``A {Versatile}
  {Technique} to {Enable} {Sub}-milli-{Kelvin} {Instrument} {Stability} for
  {Precise} {Radial} {Velocity} {Measurements}: {Tests} with the
  {Habitable}-zone {Planet} {Finder},'' {\em The Astrophysical Journal}~{\bf
  833},  175 (Dec. 2016).

\bibitem{roy_scrambling_2014}
Roy, A., Halverson, S., Mahadevan, S., and Ramsey, L.~W., ``Scrambling and
  modal noise mitigation in the {Habitable} {Zone} {Planet} {Finder} fiber
  feed,''  {\bf 9147},  91476B (July 2014).

\bibitem{halverson_efficient_2015}
Halverson, S., Roy, A., Mahadevan, S., Ramsey, L., Levi, E., Schwab, C., {Fred
  Hearty}, and MacDonald, N., ``An {Efficient}, {Compact}, and {Versatile}
  {Fiber} {Double} {Scrambler} for {High} {Precision} {Radial} {Velocity}
  {Instruments},'' {\em The Astrophysical Journal}~{\bf 806}(1),  61 (2015).

\bibitem{baudrand_modal_2001}
Baudrand, J. and Walker, G. A.~H., ``Modal {Noise} in {High}‐{Resolution},
  {Fiber}‐fed {Spectra}: {A} {Study} and {Simple} {Cure},'' {\em Publications
  of the Astronomical Society of the Pacific}~{\bf 113},  851 (July 2001).

\bibitem{mahadevan_suppression_2014}
Mahadevan, S., Halverson, S., Ramsey, L., and Venditti, N., ``Suppression of
  {Fiber} {Modal} {Noise} {Induced} {Radial} {Velocity} {Errors} for {Bright}
  {Emission}-line {Calibration} {Sources},'' {\em The Astrophysical
  Journal}~{\bf 786}(1),  18 (2014).

\bibitem{osterman_near_2012}
Osterman, S., Ycas, G.~G., Diddams, S.~A., Quinlan, F., Mahadevan, S., Ramsey,
  L., Bender, C.~F., Terrien, R., Botzer, B., Sigurddson, S., and Redman,
  S.~L., ``A near infrared frequency comb for {Y}+{J} band astronomical
  spectroscopy,'' {\em arXiv:1209.3295 [astro-ph]} ,  84501I (Sept. 2012).
\newblock arXiv: 1209.3295.

\bibitem{halverson_development_2013}
Halverson, S., Mahadevan, S., Ramsey, L.~W., Redman, S., Nave, G., Wilson,
  J.~C., Hearty, F., and Holtzman, J., ``Development of a new, precise
  near-infrared {Doppler} wavelength reference: a fiber {Fabry}-{Perot}
  interferometer,'' in [{\em Ground-based and {Airborne} {Instrumentation} for
  {Astronomy} {IV}}{\nolinebreak\hspace{0.1em}]},   {\bf 8446},  84468Q,
  International Society for Optics and Photonics (Jan. 2013).

\bibitem{halverson_development_2014}
Halverson, S., Mahadevan, S., Ramsey, L., Hearty, F., Wilson, J., Holtzman, J.,
  Redman, S., Nave, G., Nidever, D., Nelson, M., Venditti, N., Bizyaev, D., and
  Fleming, S., ``Development of {Fiber} {Fabry}-{Perot} {Interferometers} as
  {Stable} {Near}-infrared {Calibration} {Sources} for {High} {Resolution}
  {Spectrographs},'' {\em Publications of the Astronomical Society of the
  Pacific}~{\bf 126},  445 (May 2014).

\bibitem{halverson_habitable-zone_2014}
Halverson, S., Mahadevan, S., Ramsey, L., Terrien, R., Roy, A., Schwab, C.,
  Bender, C., Hearty, F., Levi, E., Osterman, S., Ycas, G., and Diddams, S.,
  ``The habitable-zone planet finder calibration system,'' in [{\em
  Ground-based and {Airborne} {Instrumentation} for {Astronomy}
  {V}}{\nolinebreak\hspace{0.1em}]},   {\bf 9147},  91477Z, International
  Society for Optics and Photonics (July 2014).

\bibitem{lee_facility_2012}
Lee, H., Hill, G.~J., Vattiat, B.~L., Smith, M.~P., and Haeuser, M., ``Facility
  calibration unit of {Hobby} {Eberly} {Telescope} wide field upgrade,'' in
  [{\em Ground-based and {Airborne} {Telescopes}
  {IV}}{\nolinebreak\hspace{0.1em}]},   {\bf 8444},  84444J, International
  Society for Optics and Photonics (Sept. 2012).

\bibitem{vattiat_design_2014}
Vattiat, B., Hill, G.~J., Lee, H., Moreira, W., Drory, N., Ramsey, J., Elliot,
  L., Landriau, M., Perry, D.~M., Savage, R., Kriel, H., Häuser, M., and
  Mangold, F., ``Design, alignment, and deployment of the {Hobby} {Eberly}
  {Telescope} prime focus instrument package,'' in [{\em Ground-based and
  {Airborne} {Instrumentation} for {Astronomy}
  {V}}{\nolinebreak\hspace{0.1em}]},   {\bf 9147},  91474J, International
  Society for Optics and Photonics (July 2014).

\bibitem{lee_lrs2:_2010}
Lee, H., Chonis, T.~S., Hill, G.~J., DePoy, D.~L., Marshall, J.~L., and
  Vattiat, B., ``{LRS}2: a new low-resolution spectrograph for the
  {Hobby}-{Eberly} {Telescope},'' in [{\em Ground-based and {Airborne}
  {Instrumentation} for {Astronomy} {III}}{\nolinebreak\hspace{0.1em}]},   {\bf
  7735},  77357H, International Society for Optics and Photonics (July 2010).

\bibitem{hill_virus:_2012}
Hill, G.~J., Tuttle, S.~E., Lee, H., Vattiat, B.~L., Cornell, M.~E., DePoy,
  D.~L., Drory, N., Fabricius, M.~H., Kelz, A., Marshall, J.~L., Murphy, J.~D.,
  Prochaska, T., Allen, R.~D., Bender, R., Blanc, G., Chonis, T., Dalton, G.,
  Gebhardt, K., Good, J., Haynes, D., Jahn, T., MacQueen, P.~J., Rafal, M.~D.,
  Roth, M.~M., Savage, R.~D., and Snigula, J., ``{VIRUS}: production of a
  massively replicated 33k fiber integral field spectrograph for the upgraded
  {Hobby}-{Eberly} {Telescope},'' in [{\em Ground-based and {Airborne}
  {Instrumentation} for {Astronomy} {IV}}{\nolinebreak\hspace{0.1em}]},   {\bf
  8446},  84460N, International Society for Optics and Photonics (Sept. 2012).

\bibitem{seifahrt_development_2016}
Seifahrt, A., Bean, J.~L., Stürmer, J., Gers, L., Grobler, D.~S., Reed, T.,
  and Jones, D.~J., ``Development and construction of {MAROON}-{X},'' in [{\em
  Ground-based and {Airborne} {Instrumentation} for {Astronomy}
  {VI}}{\nolinebreak\hspace{0.1em}]},   {\bf 9908},  990818, International
  Society for Optics and Photonics (Aug. 2016).

\bibitem{ramsey_focal_1988}
Ramsey, L.~W., ``Focal ratio degradation in optical fibers of astronomical
  interest,''  {\bf 3},  26--39 (1988).

\bibitem{mahadevan_inexpensive_2008}
Mahadevan, S., Ge, J., Fleming, S.~W., Wan, X., DeWitt, C., van Eyken, J.~C.,
  and McDavitt, D., ``An {Inexpensive} {Field}-{Widened} {Monolithic}
  {Michelson} {Interferometer} for {Precision} {Radial} {Velocity}
  {Measurements},'' {\em Publications of the Astronomical Society of the
  Pacific}~{\bf 120},  1001--1015 (Sept. 2008).
\newblock arXiv: 0809.3721.

\bibitem{chaffee_astronomical_1976}
Chaffee, Jr., F.~H. and Schroeder, D.~J., ``Astronomical applications of
  echelle spectroscopy,'' {\em Annual Review of Astronomy and
  Astrophysics}~{\bf 14},  23--42 (1976).

\end{thebibliography}
\bibliographystyle{spiebib} % makes bibtex use spiebib.bst

\end{document}